\documentclass[preprintnumbers,jcp,showpacs,superscriptaddress]{revtex4-1}

\usepackage{palatino}
\usepackage[latin1]{inputenc}
\usepackage{epsf}
\usepackage{amsmath,amssymb}
\usepackage{latexsym}
\usepackage{calc}
\usepackage{color}
\usepackage{epsfig}

\def\dulR{{\underline{\underline{\bf R}}}}
\def\dulr{{\underline{\underline{\bf r}}}}
\def\bA{{\bf A}}

\begin{document}
\title{The exact forces on classical nuclei in non-adiabatic charge transfer}

\author{Federica Agostini$^a$, Ali Abedi$^{a}$, Yasumitsu Suzuki$^{a}$, Seung Kyu Min$^{a}$, Neepa T. Maitra$^b$ and E. K. U. Gross$^{a}$\\
\vspace{6pt} $^{a}${\em{Max-Planck-Institut of Microstructure Physics, Weinberg 2, D-06120 Halle, Germany}}\\
\vspace{6pt} $^{b}${\em{Department of Physics and Astronomy, Hunter College and the Graduate Center of the City University of New York, 695 Park Avenue, New York, New York 10065, United States}}\\
\vspace{6pt} $^{}$}
\date{\today}
\begin{abstract}
The 
decomposition of electronic and nuclear motion presented
in~[A. Abedi, N. T. Maitra, and E. K. U. Gross, Phys. Rev. Lett. 105,
  123002 (2010)] yields a time-dependent potential that drives
the nuclear motion and fully accounts for the coupling to the
electronic subsystem. 
Here we show that  propagation of an ensemble of independent classical nuclear
trajectories  on this exact potential yields
dynamics that are essentially indistinguishable from the exact quantum
dynamics for a model non-adiabatic charge transfer problem. 
We point out the importance of step and bump features in the exact potential that are critical in obtaining 
the correct splitting of the quasiclassical nuclear wave packet
in space after it passes
through an
avoided crossing between two Born-Oppenheimer surfaces, and analyze their structure. 
Lastly, an analysis of the exact potentials in the context of  trajectory surface hopping procedure is presented, including preliminary investigations of velocity-adjustment, and the force-induced decoherence effect. 
\end{abstract}
\maketitle 

\section{Introduction}

The Born-Oppenheimer~\cite{BO} (BO), or adiabatic, approximation is
amongst the most fundamental approximations in physics and chemistry, and is at the basis of our understanding of the coupled
electron-nuclear dynamics in molecules and solids. At the heart of the
BO approximation lies the assumption that, due to the high
ratio between nuclear and electronic masses, the electrons move much faster than the nuclei and adjust
instantaneously to the positions of the slower-moving nuclei.   A molecule or solid is
then viewed as a set of nuclei moving on a single potential energy
surface (PES) generated by the electrons in a given
eigenstate. Moreover, the fundamental constructs of the BO approach, the BOPESs, form the launching ground for methods that describe processes beyond the
adiabatic regime  where the  BO approximation itself
fails.
Prominent examples of such electronic non-adiabatic processes appear throughout physics, chemistry and biology, for example
vision~\cite{cerulloN2010, schultenBJ2009, ishidaJPCB2012},
photo-synthesis~\cite{tapaviczaPCCP2011, flemingN2005},
photo-voltaics~\cite{rozziNC2013, silvaNM2013, jailaubekovNM2013},
proton-transfer/hydrogen storage~\cite{sobolewski, varella,
  hammes-schiffer, marx}.

The standard approaches to describe non-adiabatic molecular processes
are in terms of coupled BOPESs and transitions between the corresponding adiabatic electronic states induced by the nuclear
motion.
In the Born-Huang expansion, the exact solution of the time-dependent Schr\"odinger equation (TDSE)
is expanded in the complete set of BO electronic states, leading to a nuclear wave packet with
contributions on several BOPESs that undergo transitions in the
regions of strong non-adiabatic coupling. This formally exact approach
is hard to use in practice because of the high computational cost of
describing the nuclear time evolution quantum mechanically.  Approximate methods
that retain a quantum description of nuclei have been successful in
some applications (e.g. multiple-spawning~\cite{martinezCPL1996, martinezJPC1996, martinezACR2006},  multiconfiguration time-dependent {Hartree}~\cite{cederbaumCPL1990, burghardtJCP1999, thossCM2004}, or non-adiabatic Bohmian dynamics~\cite{tavernelliPCCP2011, tavernelliPRA2013} methods), but are
still limited due to the computational effort required.
Therefore, approaches that involve a classical or semi-classical
description of the nuclear motion that is non-adiabatically coupled to
the (quantum mechanical) electrons become methods of choice due to
their applicability to large systems. Ubiquitous are the Ehrenfest and
surface-hopping methods.  In developing a mixed quantum-classical
approach, however, one faces two challenging questions: What are the true
classical forces acting on the nuclear subsystem? How is the coupling
between  the nuclear classical trajectories and the electronic subsystem
defined? Despite extensive studies of both new and routinely used
methods~\cite{pechukasPR1969_1, pechukasPR1969_2, ehrenfest, TSH, TSH_1990, wyattJCP2001, wyattJCP2002, parlantJPCA2007, tullyJCP2000, tannorJCP2012, garashchukJCP2003, prezhdoPRL2001, shenvi-subotnikJCP2011, landryJCP2013_2, kapral-ciccotti, kapralARPC2006, io, gevaJCP2004, schuetteJCP2001, martensJCP1997, bonellaCP2001, cokerJCP2012, millerJCP1997, thossPRL1997, millerJCP2007, millerJPCA2009, thossARPC2004, hermannARPC1994, hermanJCP2005, tavernelliPCCP2011, tavernelliPRA2013, truhlarFD2004, truhlarACR2006, burghardtJCP2011, jangJCP2012, vanicekJCP2012, takatsukaPRA2010},
there is still no clear answer to the aforementioned questions.

In recent work we have resolved the question of what force drives the
nuclei~\cite{steps,long_steps,mqc} based on the novel perspective offered
by the exact decomposition of electronic and nuclear motion of
Refs.~\cite{AMG,AMG2}.  The full molecular wave function is represented
as a single product of a purely nuclear wave function and an
electronic factor that parametrically depends on the nuclear
coordinates~\cite{AMG,AMG2}.  We have shown that in this framework,
the nuclear dynamics is governed by a TDSE that contains a time
dependent potential energy surface (TDPES) and a time-dependent vector
potential. These potentials are rigorous concepts and
provide the exact \textit{driving forces} of the nuclear evolution.
Refs.~\cite{steps, long_steps} showed that step-like features in these
exact potentials mediate coupling between piecewise Born-Oppenheimer
surfaces.  Ref.~\cite{long_steps} demonstrated that a classical
nuclear trajectory evolving on the exact TDPES can capture accurately
averaged observables, such as the mean nuclear position, but fails to capture the splitting of the nuclear wave packet
that occurs after passing through an avoided crossing between adiabatic
surfaces. Refs.~\cite{mqc, long_mqc} explored a self-consistent mixed quantum-classical scheme, treating the nuclei classically in both the electronic and nuclear equations of motion, finding reasonable accuracy for averaged observables but again no wave packet splitting.

In the present paper, we analyze in detail  a model
non-adiabatic charge transfer process within the exact factorization
framework when the nuclei are evolved classically.  We consider a
one-dimensional (1D) model simple enough to allow for an exact
solution of the full TDSE, while at the same time exhibiting
characteristic features associated with non-adiabatic dynamics.  

Our main results are threefold. First, we show that, once an {\it
  ensemble} of independent classical trajectories is evolved on the
exact TDPES, splitting of the nuclear wave packet characterizing a
non-adiabatic event can be captured.  This therefore overcomes the
limitations of the single trajectory dynamics of
Ref.~\cite{long_steps}. In fact, it turns out that the resulting
quasiclassical evolution is essentially identical to the exact quantum
nuclear dynamics.  The representation of the quantum nuclear wave
packet as an ensemble of independent trajectories is the only
approximation we make: the forces acting on the classical nuclei are
obtained from the exact potentials acting on the nuclear subsystem,
containing, in the parlance of mixed quantum-classical methods, the
exact ``quantum electronic back-reaction''.  The features of the time
dependent potential responsible for the splitting, and in general for
the evolution of the nuclei, are analyzed in detail and compared with
the standard picture in terms of static BOPESs.  Importantly, we
highlight the significance of the gauge-dependent contribution to the
exact TDPES: without this term the quasiclassical dynamics is poor. In
previous work~\cite{steps,long_steps} the structure of the
gauge-independent (GI) term has been analyzed for this problem, and it
was observed that the gauge-dependent (GD) term appears to largely
cancel step structures in the GI term.  This brings us to our second
main result: we analyze, with the help of some semiclassical
arguments, the structure of the GD term, and explain why it
consists of piecewise flat segments joined by steps which almost, but
not completely, cancel the steps in the GI term.  Our third main
result concerns the question of what mixed quantum-classical dynamics
based on the exact TDPES can tell us about standard approaches such as
trajectory surface-hopping (TSH). In particular, how does the exact
force on a trajectory evolving on the TDPES compare to the force it
experiences in TSH? We find several aspects that are somewhat qualitatively in
common: after passing through an avoided crossing region, the exact
TDPES tracks one BOPES or the other piecewise in space; moreover, it
displays an energy adjustment between the surfaces, which can been (gauge-)transformed to a kinetic energy contribution, reminiscent of the velocity-adjustment in TSH. We show that the exact TDPES sheds light on the notorious
problem of over-coherence in TSH~\cite{shenvi-subotnikJCP2011, landryJCP2013_2, rosskyJCP1995, prezhdoJCP2012, prezhdoJCP2014, tavernelliJCP2013, truhlarFD2004, truhlarACR2006, makriCPL2014, kapralJCP2008, persicoJCP2007} 
 by comparing the electronic density-matrix associated to trajectories in each case. The force
resulting from the step features in the TDPES appear to be intimately
related to creating decoherence, lacking in TSH.

The larger context in which these results stand is the eventual development of a new mixed quantum-classical approach to non-adiabatic dynamics based on potentials arising out of the exact factorization of the electron-nuclear wave function. In the present paper we do not derive an
  algorithm, but we show the features of the classical force that
  such a numerical scheme should reproduce. In realistic systems the exact forces acting on the nuclei will need to be approximated, and these
approximations should build in the features uncovered in the present work in order to yield accurate dynamics. Further, it is extremely instructive to present our analysis in
  comparison to other methods, as this comparison gives an idea of the
  real possibilities of our procedure in perspective to what has been
  already proposed in the literature. Moreover, being aware of the
  difficulties in interpreting the new formalism, we try to put our
  new findings in the light of \textit{old} concepts, in order to help
  the reader understanding the potentiality of the method. We devote Appendix A to a detailed discussion about the connections between the present trajectory-based
exact factorization approach and Bohmian dynamics; the point is to stress that despite both theories
involve trajectories propagating on a time-dependent potential, the present theory is completely independent of the Bohmian approach.

The paper is organized as follows. A general introduction to the exact
decomposition of electronic and nuclear motion is given in
Section~\ref{sec: background}, followed by a discussion about gauge
conditions.  The model system is introduced in Section~\ref{sec:
  model} and its dynamics, both fully exact quantum and with the
classical approximation for nuclei, are analyzed in Section~\ref{sec:
  cl vs. qm}, along with the averaged nuclear observables (Section~\ref{sec: observables}). Section~\ref{sec: gd potentials}
then focusses on the gauge-dependent contribution to the potential,
analyzing its structure in the TDPES (Section~\ref{sec: gd}), and then
gauge-transforming it to a vector potential (Section~\ref{sec: A}) for
a different perspective of its effect on the dynamics.
Section~\ref{sec: decoherence} is devoted to the topic of decoherence. Our conclusions and perspectives are summarized in Section~\ref{sec: conclusions}.

\section{Exact decomposition of the electronic and nuclear motion}\label{sec: background}
The non-relativistic Hamiltonian describing a system of interacting electrons and nuclei, in the absence of a time-dependent external field, is
\begin{equation}\label{eqn: hamiltonian}
 \hat H = \hat T_n+\hat H_{BO},
\end{equation}
where $\hat T_n$ is the nuclear kinetic energy operator and 
\begin{equation}\label{eqn: boe}
\hat{H}_{BO}(\dulr,\dulR) = \hat{T}_e(\dulr) + \hat{W}_{ee}(\dulr) + \hat{V}_{en}(\dulr,\dulR) +
\hat{W}_{nn}(\dulR)
\end{equation}
is the standard BO electronic Hamiltonian, with electronic kinetic energy $\hat{T}_e(\dulr)$, and interaction potentials  $\hat{W}_{ee}(\dulr)$  for electron-electron, $\hat{W}_{nn}(\dulR)$ for nucleus-nucleus, and  $\hat{V}_{en}(\dulr,\dulR)$ for electron-nucleus.
The symbols $\dulr$ and $\dulR$ are used to collectively indicate the 
coordinates of $N_{e}$ electrons and $N_n$ nuclei, respectively.

It has been proved~\cite{AMG,AMG2}, that the full time-dependent electron-nuclear wave function, $\Psi(\dulr,\dulR,t)$, that is the solution of the TDSE,
\begin{equation}\label{eqn: tdse}
 \hat H\Psi(\dulr,\dulR,t)=i\hbar\partial_t \Psi(\dulr,\dulR,t),
\end{equation}
can be exactly factorized to the correlated product
\begin{equation}\label{eqn: factorization}
 \Psi(\dulr,\dulR,t)=\chi(\dulR,t)\Phi_\dulR(\dulr,t)
\end{equation}
 where 
\begin{equation}
 \int d\dulr \left|\Phi_\dulR(\dulr,t)\right|^2 = 1 \quad\forall\,\,\dulR,t.
\label{eq:PNC}
\end{equation}
Here $\chi(\dulR,t)$ is the nuclear wave function and $\Phi_\dulR(\dulr,t)$ is the electronic wave function  which
parametrically depends on the nuclear positions and satisfies the partial normalization condition (PNC) expressed in Eq.~(\ref{eq:PNC}). 
The PNC guarantees the interpretation of $|\chi(\dulR,t)|^2$ as the
probability of finding the nuclear configuration $\dulR$ at time $t$, and of
$|\Phi_\dulR(\dulr,t)|^2$ itself as the conditional
probability of finding the electronic configuration $\dulr$ at time $t$ for nuclear configuration $\dulR$.  Further, the PNC makes the factorization~(\ref{eqn:
  factorization}) unique up to within a $(\dulR,t)$-dependent gauge
transformation,
\begin{equation}\label{eqn: gauge}
 \begin{array}{rcl}
  \chi(\dulR,t)\rightarrow\tilde\chi(\dulR,t)&=&e^{-\frac{i}{\hbar}\theta(\dulR,t)}\chi(\dulR,t) \\
  \Phi_\dulR(\dulr,t)\rightarrow\tilde\Phi_\dulR(\dulr,t)&=&e^{\frac{i}{\hbar}\theta(\dulR,t)}\Phi_\dulR(\dulr,t),
 \end{array}
\end{equation}
 where $\theta(\dulR,t)$ is some real function of the nuclear coordinates and time. 

The stationary variations~\cite{frenkel} of the quantum mechanical action with respect to $\Phi_\dulR(\dulr,t)$ and $\chi(\dulR,t)$ lead to the derivation of the following equations of motion
\begin{eqnarray}
 \left(\hat H_{el}(\dulr,\dulR)-\epsilon(\dulR,t)\right)
 \Phi_{\dulR}(\dulr,t)&=&i\hbar\partial_t \Phi_{\dulR}(\dulr,t)\label{eqn: exact electronic eqn} \\ 
 \hat H_n(\dulR,t)\chi(\dulR,t)&=&i\hbar\partial_t \chi(\dulR,t), \label{eqn: exact nuclear eqn}
\end{eqnarray}
where the PNC is inserted by means of Lagrange multipliers~\cite{alonsoJCP2013, AMG2013}. Here, the electronic and nuclear Hamiltonians are defined as
\begin{equation}\label{eqn: electronic hamiltonian}
\hat H_{el}(\dulr,\dulR)=\hat{H}_{BO}(\dulr,\dulR)+\hat U_{en}^{coup}[\Phi_\dulR,\chi]
\end{equation}
and
\begin{equation}\label{eqn: nuclear hamiltonian}
\hat H_n(\dulR,t) = \sum_{\nu=1}^{N_n} \frac{\left[-i\hbar\nabla_\nu+\bA_\nu(\dulR,t)\right]^2}{2M_\nu} + \epsilon(\dulR,t),
\end{equation}
respectively, with ``electron-nuclear coupling operator''
\begin{align}
\hat U_{en}^{coup}&[\Phi_\dulR,\chi]=\sum_{\nu=1}^{N_n}\frac{1}{M_\nu}\left[
 \frac{\left[-i\hbar\nabla_\nu-\bA_\nu(\dulR,t)\right]^2}{2} \right.\label{eqn: enco} \\
& \left.+\left(\frac{-i\hbar\nabla_\nu\chi}{\chi}+\bA_\nu(\dulR,t)\right)
 \left(-i\hbar\nabla_\nu-\bA_{\nu}(\dulR,t)\right)\right].\nonumber
\end{align}
The potentials in the theory are the scalar TDPES, $\epsilon(\dulR,t)$, implicitly defined by Eq.~(\ref{eqn: exact electronic eqn}) as
\begin{equation}\label{eqn: tdpes}
 \epsilon(\dulR,t)=\left\langle\Phi_\dulR(t)\right|\hat{H}_{BO}+\hat U_{en}^{coup}-i\hbar\partial_t\left|
 \Phi_\dulR(t)\right\rangle_\dulr,
\end{equation}
and the time-dependent vector potential, $\bA_{\nu}\left(\dulR,t\right)$, defined as
\begin{equation}\label{eqn: vector potential}
 \bA_{\nu}\left(\dulR,t\right) = \left\langle\Phi_\dulR(t)\right|-i\hbar\nabla_\nu\left.\Phi_\dulR(t)
 \right\rangle_\dulr\,.
\end{equation}
The symbol $\left\langle\,\,\cdot\,\,\right\rangle_\dulr$ indicates an integration over electronic coordinates only. The $\dulr$-de\-pen\-den\-ce of the electronic wave function is dropped only for formal consistency between the left-hand-side of Eqs.~(\ref{eqn: tdpes}) and~(\ref{eqn: vector potential}), that do not depend on $\dulr$, and the right-hand-side, where the $\dulr$-dependence is integrated out. The electronic wave function itself remains a parametric function of $\dulR$ and a function of $\dulr,t$.
Under the gauge transformation~(\ref{eqn: gauge}),
the scalar potential and the vector potential transform as  
\begin{eqnarray}
\tilde{\epsilon}(\dulR,t) &=& \epsilon(\dulR,t)+\partial_t\theta(\dulR,t)\label{eqn: transformation of epsilon} \\
\tilde{\bf A}_{\nu}(\dulR,t) &=& {\bf A}_{\nu}(\dulR,t)+\nabla_\nu\theta(\dulR,t)\,.\label{eqn: transformation of A}
\end{eqnarray}
In Eqs.~(\ref{eqn: exact electronic eqn}) and~(\ref{eqn: exact nuclear
  eqn}), $\hat U_{en}^{coup}[\Phi_\dulR,\chi]$, $\epsilon(\dulR,t)$
and $\bA_{\nu}\left(\dulR,t\right)$ are responsible for the coupling
between electrons and nuclei in a formally exact way. It is worth
noting that the electron-nuclear coupling operator, $\hat
U_{en}^{coup}[\Phi_\dulR,\chi]$, in the electronic equation~(\ref{eqn:
  exact electronic eqn}), depends on the nuclear wave function and
acts on the parametric dependence of $\Phi_\dulR(\dulr,t)$ as a
differential operator. This ``pseudo-operator'' includes the coupling
to the nuclear subsystem beyond the parametric dependence in the BO
Hamiltonian $\hat H_{BO}(\dulr,\dulR)$. It is worth noting that the shape of TDPES at time $t$ depends on the choice of initial electronic state $\Phi_\dulR(\dulr,t=0)$. Here, we have chosen a BO electronic state as the initial electronic wave function to compare the TDPES with the standard picture of static BOPESs.

The nuclear equation~(\ref{eqn: exact nuclear eqn}) has the
particularly appealing form of a Schr\"odinger equation that contains
a time-dependent vector potential~(\ref{eqn: vector potential}) and a
time-dependent scalar potential~(\ref{eqn: tdpes}) that govern the
nuclear dynamics and yield the nuclear wave function. These potentials are uniquely determined up to within the gauge
transformation, given in Eqs.~(\ref{eqn: transformation of epsilon})
and~(\ref{eqn: transformation of A}). As expected, the nuclear
Hamiltonian in Eq.~(\ref{eqn: exact nuclear eqn}) is form-invariant
under such transformations. $\chi(\dulR,t)$ is interpreted as the
nuclear wave function since it leads to an $N$-body nuclear density,
\begin{equation}\label{eqn: exact nuclear density}
\Gamma(\dulR,t)=\vert\chi(\dulR,t)\vert^2,
\end{equation}
 and an $N$-body current-density, 
\begin{equation} 
 {\bf J}_\nu(\dulR,t)=\frac{\Big[\mbox{Im}(\chi^*(\dulR,t)\nabla_\nu\chi(\dulR,t))+ \Gamma(\dulR,t){\bf A}_\nu(\dulR,t)\Big]}{M_\nu},
\end{equation}
which reproduce the true nuclear $N$-body density and current-density~\cite{AMG2} obtained from the full wave function
$\Psi(\dulr,\dulR,t)$. The uniqueness of $\epsilon(\dulR,t)$ and
$\bA_{\nu}(\dulR,t)$ can be straightforwardly proved by following the
steps of the current-density version~\cite{Ghosh-Dhara} of the
Runge-Gross theorem~\cite{RGT}, or by referring to the theorems proved in Ref.~\cite{AMG}. 

\subsection{The choice of the gauge}
The results of any calculation do not depend on the choice of gauge in
Eqs.~(\ref{eqn: gauge}). The form of the TDPES and vector potential do depend on the choice, but together their effect on the dynamics, on all observables, electronic populations etc, is gauge-independent. 
It is instructive to decompose the TDPES into gauge-invariant (GI) and
gauge-dependent (GD) constituents,
\begin{equation}
 \epsilon(\dulR,t)=\epsilon_{GI}(\dulR,t)+\epsilon_{GD}(\dulR,t),
\end{equation}
where
\begin{align}
 \epsilon_{GI}&(\dulR,t)=\left\langle\Phi_\dulR(t)\right|\hat{H}_{BO}\left|\Phi_\dulR(t)\right\rangle_\dulr \label{eqn: gi tdpes} \\
 &+\sum_{\nu=1}^{N_n}\bigg(\frac{\hbar^2}{2M_\nu}\left\langle\nabla_\nu\Phi_\dulR(t)|\nabla_\nu\Phi_\dulR(t)
 \right\rangle_\dulr-\frac{\bA^2_\nu(\dulR,t)}{2M_\nu}\bigg),\nonumber
\end{align}
with the second term on the right-hand-side obtained from the action of the electron-nuclear coupling operator in Eq.~(\ref{eqn: enco}) 
on the electronic wave function, and
\begin{equation}\label{eqn: gd tdpes}
 \epsilon_{GD}(\dulR,t)=\left\langle\Phi_\dulR(t)\right|-i\hbar\partial_t\left|\Phi_\dulR(t)\right\rangle_\dulr.
\end{equation}
The GI part of the TDPES, $\epsilon_{GI}$, is invariant under the
gauge transformation~(\ref{eqn: gauge}):
$\tilde{\epsilon}_{GI}(\dulR,t)=\epsilon_{GI}(\dulR,t)$. The GD part,
on the other hand, transforms as
$\tilde{\epsilon}_{GD}(\dulR,t)=\epsilon_{GD}(\dulR,t)+\partial_t\theta(\dulR,t)$.

For purposes of our analysis, to help understand the exact
potentials in coupled electron-ion dynamics, and their comparison to
traditional methods, we will find two gauges are particularly useful.
First, we note that the model system we study is in 1D, so
the vector potential can be gauged away. In general
three-dimensional cases where the vector potential is curl-free, also the
gauge may be chosen where the vector potential is zero
$\bA_\nu(\dulR,t)\equiv0$. Whether and under which conditions
$\mbox{curl}\,\bA_\nu(\dulR,t)=0$ is, at the moment, under
investigations~\cite{CI_MAG}.

The first gauge we will consider is one where the $\bA_\nu(\dulR,t)\equiv0$, and
therefore  the entire electronic back-reaction is contained in the
TDPES. To determine this, consider first that the nuclear wave function is fully determined by its modulus and phase, according to $\chi(\dulR,t)=|\chi(\dulR,t)|e^{iS(\dulR,t)/\hbar}$. The condition
\begin{equation}
 \left|\chi(\dulR,t)\right|=\sqrt{\int d\dulr \left|\Psi(\dulr,\dulR,t)\right|^2},
\label{eq:modchi}
\end{equation}
on the modulus, automatically satisfies the request that the exact nuclear density calculated from $\Psi$ can be also obtained directly from $\chi$. On the other hand, from Eqs.~(\ref{eqn: gauge}), we notice that the phase $S(\dulR,t)$ of the nuclear wave function is related to the choice of gauge. 
If we 
impose 
\begin{equation}
\label{eq:nphase}
 S(R,t)=\int^R dR'\frac{\mbox{Im}\left\langle\Psi(t)\right|\left.\partial_{R'}\Psi(t)\right\rangle_r}{\left|\chi(R',t)\right|^{2}}\;,
\end{equation}
we find $A(R,t)=0$ as we will now show. 
Note that here we dropped the bold double-underlined symbols, in order to represent electronic and nuclear coordinates in 1D. Henceforth, the old symbols will be used whenever our statements have general validity and the new symbols will be used for the 1D case only. It can be easily proved that Eq.~(\ref{eq:nphase}) results in a vector-potential-free gauge. To do this, we
 insert the factorization~(\ref{eqn: factorization}) in the expression of the vector potential~(\ref{eqn: vector potential}), obtaining~\cite{AMG} a relation between the vector potential itself and the nuclear velocity field
\begin{equation}\label{eqn: vector-exact}
 \bA_\nu(\dulR,t) = \frac{\mbox{Im}\left\langle\Psi(t)\right|\left.\nabla_\nu
 \Psi(t)\right\rangle_{\dulr}}{\left|\chi(\dulR,t)\right|^{2}} - \nabla_\nu S(\dulR,t),
\end{equation}
that in 1D reads $A(R,t) = \mbox{Im}\left\langle\Psi(t)\right|\left.\partial_R
 \Psi(t)\right\rangle_{r}/\left|\chi(R,t)\right|^{2} - \partial_R S(R,t)$. 
 Imposing here $A(R,t)=0$ leads to Eq.~(\ref{eq:nphase}), which defines the phase of the nuclear wave function.
We then obtain the TDPES, $\epsilon(\dulR,t)$ from Eq.~(\ref{eqn: tdpes}), by explicitly calculating the electronic wave function, $\Phi_\dulR(\dulr,t)=\Psi(\dulr,\dulR,t)/\chi(\dulR,t)$. Alternatively, we may invert the nuclear equation~(\ref{eqn: exact nuclear eqn}) to find the TDPES.

We have used this vector-potential-free gauge to perform the classical
calculations. The TDPES alone determines the time evolution of
$\chi(\dulR,t)$ and has both GI and GD
components, Eqs.~(\ref{eqn: gi tdpes}) and~(\ref{eqn: gd tdpes})
above.  In Section~\ref{sec: gd}, we will discuss the characteristic
features of $\epsilon_{GI}$ and $\epsilon_{GD}$, with particular
attention to the latter. The former has been extensively analyzed before~\cite{steps,long_steps} and the main results will be
briefly recalled.

The second gauge which we will find instructive to study
(Section~\ref{sec: A}) is one where we instead transform the
  gauge-dependent part of the TDPES, $\epsilon_{GD}$, into a vector
  potential. In some sense, this gauge makes a more direct comparison
  with the TSH methods, and we will see hints of ``velocity
  adjustment'' used in TSH, appearing in the exact vector
  potential.

\subsection{Comparison with Born-Huang expansion}
A major theme of this work and the recent papers~\cite{AMG2, steps,
  long_steps} is the relation of the exact TDPES to the BOPESs.
 Therefore, we introduce here the
BO electronic states, $\varphi_{\dulR}^{(l)}(\dulr)$, and BOPESs,
$\epsilon_{BO}^{(l)}(\dulR)$, which are the normalized eigenstates and
eigenvalues of the BO electronic Hamiltonian~(\ref{eqn: boe}),
respectively. If the full wave function is expanded in this basis,
\begin{equation}\label{eqn: expansion of Psi}
 \Psi(\dulr,\dulR,t)=\sum_l F_l(\dulR,t)\varphi_\dulR^{(l)}(\dulr),
\end{equation} 
then the nuclear density may be written as  
\begin{equation}\label{eqn: chi and Fl}
 \left|\chi(\dulR,t)\right|^2 = \sum_{l}\left|F_l(\dulR,t)\right|^2.
\end{equation}
This relation is obtained by integrating the squared modulus of Eq.~(\ref{eqn: expansion of Psi}) over the electronic 
coordinates. The exact electronic wave function may also be expanded in terms of the BO states,
\begin{equation}\label{eqn: expansion of Phi}
 \Phi_\dulR(\dulr,t)=\sum_l C_l(\dulR,t)\varphi_\dulR^{(l)}(\dulr).
\end{equation} 
The expansion coefficients in Eqs.~(\ref{eqn: expansion of Psi}) and~(\ref{eqn: expansion of Phi}) are related,
\begin{equation}\label{eqn: relation coefficients}
 F_l(\dulR,t)= C_l(\dulR,t)\chi(\dulR,t),
\end{equation}
by virtue of the factorization~(\ref{eqn: factorization}). The PNC then reads
\begin{equation}\label{eqn: PNC on BO}
 \sum_l\left|C_l(\dulR,t)\right|^2=1\quad\forall\,\,\dulR,t.
\end{equation}
We point out that even in the case where the nuclear wave packet splits onto more than one BOPES  the full wave 
function is still a single product: the nuclear wave function has 
 contributions 
(projections) on different BOPES while the electronic wave function is a 
linear combination of different adiabatic states, but still we may write
\begin{align}
\Psi(\dulr,&\dulR,t)=\nonumber \\
&\left(e^{\frac{i}{\hbar}S(\dulR,t)}\sqrt{\sum_{l}|F_l(\dulR,t)|^2}\right)\left(\sum_kC_k(\dulR,t)\varphi_\dulR^{(k)}(\dulr)\right)
\end{align}
where the first term in parenthesis is $\chi(\dulR,t)$, using Eqs.~(\ref{eq:nphase}) and~(\ref{eqn: chi and Fl}), and the second term in parenthesis is $\Phi_\dulR(\dulr,t)$, using Eq.~(\ref{eqn: expansion of Phi}). Ref.~\cite{steps} provides a visualization of the electronic wave function.

\section{Non-adiabatic electron transfer}\label{sec: model}
\begin{figure}[h!]
 \centering
 \includegraphics*[width=.6\textwidth]{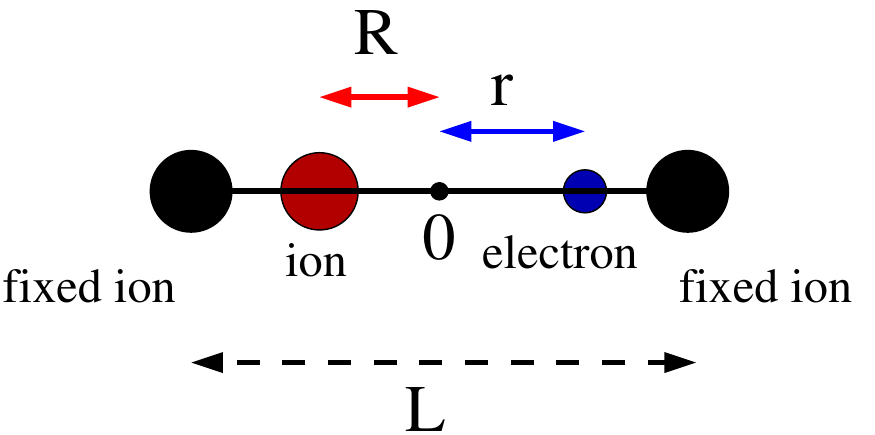}
 \caption{Schematic representation of the model system described by the Hamiltonian~(\ref{eqn: metiu-hamiltonian}). $R$ and $r$ indicate the coordinates of the moving ion and electron, respectively, in one dimension. $L$ is the distance between the fixed ions.}
 \label{fig: metiu model}
\end{figure}

\begin{figure}[h!]
 \begin{center}
 \includegraphics[angle=270,width=.85\textwidth]{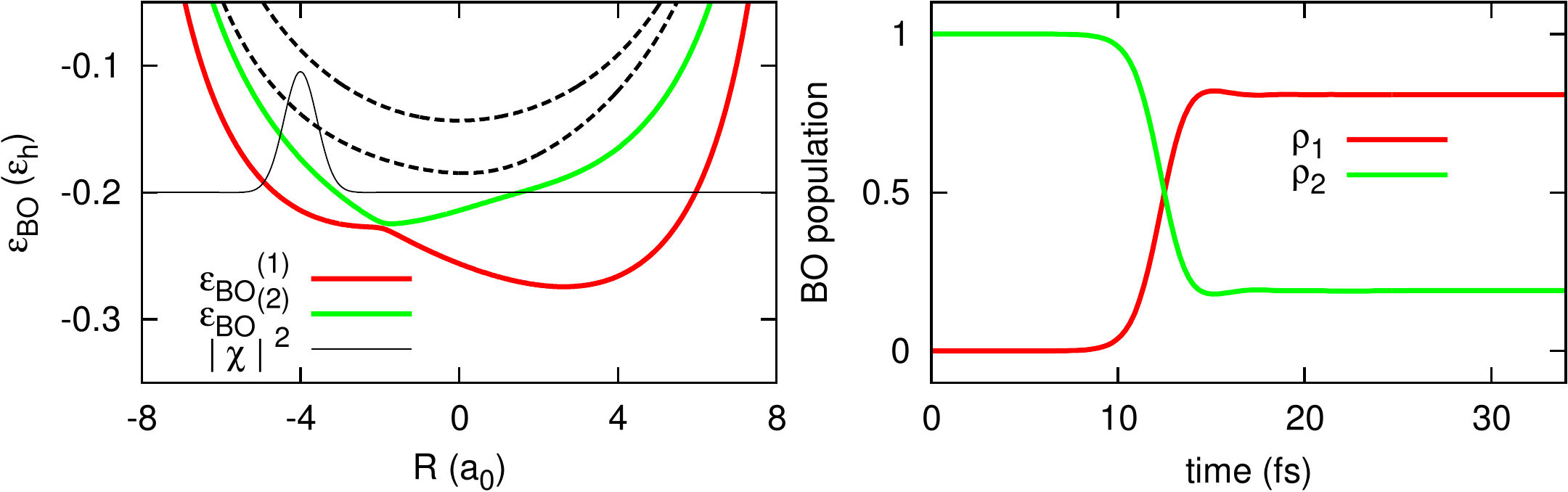}
 \caption{Left: lowest four BO surfaces, as functions of the nuclear coordinate. The first (red line) and second (green line) surfaces will be considered in the actual calculations that follow, the third and fourth (dashed black lines) are shown for reference. The squared modulus (reduced by ten times and rigidly shifted in order to superimpose it on the energy curves) of the initial nuclear wave packet is also shown (thin black line). Right: populations of the BO states as functions of time. The strong non-adiabatic nature of the model is underlined by the population exchange at the crossing of the coupling region.}
 \label{fig: BO}
 \end{center}
\end{figure}

We study here the 1D Shin-Metiu model~\cite{MM} for non-adiabatic electron transfer. The system consists of three ions and a single electron, as depicted in Fig.~\ref{fig: metiu model}.
Two ions are fixed at a distance of $L=19.0$~$a_0$, the third ion and the electron are free to move in one dimension along the line joining the two fixed ions. The Hamiltonian of this system reads
\begin{align}
 \hat{H}(r,R)= -\frac{1}{2}\frac{\partial^2}{\partial r^2}-\frac{1}{2M}\frac{\partial^2}{\partial R^2}  +
 \frac{1}{\left|\frac{L}{2}-R\right|}+\frac{1}{\left|\frac{L}{2} + R\right|}\label{eqn: metiu-hamiltonian}\\
 -\frac{\mathrm{erf}\left(\frac{\left|R-r\right|}{R_f}\right)}{\left|R - r\right|}
 -\frac{\mathrm{erf}\left(\frac{\left|r-\frac{L}{2}\right|}{R_r}\right)}{\left|r-\frac{L}{2}\right|}
 -\frac{\mathrm{erf}\left(\frac{\left|r+\frac{L}{2}\right|}{R_l}\right)}{\left|r+\frac{L}{2}\right|}.\nonumber
\end{align}
Here, the symbols $\dulr$ and $\dulR$ are replaced by $r$ and $R$, the
coordinates of the electron and the movable ion measured from the
center of the two fixed ions. The ionic mass is chosen as $M=1836$,
the proton mass, whereas the other parameters are tuned in order to make the 
system essentially a two-electronic-state model. We present here the results obtained by choosing
$R_f=5.0$~$a_0$, $R_l=3.1$~$a_0$ and $R_r=4.0$~$a_0$ such that the
first BOPES, $\epsilon^{(1)}_{BO}$, is strongly coupled to the second
BOPES, $\epsilon^{(2)}_{BO}$, around the avoided crossing at
$R_{ac}=-1.90~a_0$ and there is a weak coupling to the rest of the
surfaces. 
The BO surfaces are shown in Fig.~\ref{fig: BO}
(left panel). The analysis developed in the following sections is 
very general and does not depend on the strength of the non-adiabatic 
coupling, similar to 
Ref.~\cite{long_steps}.
 We show here results from a strong-coupling set of parameters, and the conclusions we draw here extend also to the weaker-coupling case (not shown). 

We study the time evolution of this system by choosing the initial
wave function as the product of a real-valued normalized Gaussian wave
packet, centered at $R_c=-4.0$~$a_0$ with variance
$\sigma=1/\sqrt{2.85}$~$a_0$ (thin black line in Fig.~\ref{fig: BO}),
and the second BO electronic state, $\varphi_{R}^{(2)}(r)$. To obtain
the TDPES, we first solve the TDSE~(\ref{eqn: tdse}) for the complete
system, with Hamiltonian~(\ref{eqn: metiu-hamiltonian}), and obtain
the full wave function, $\Psi(r,R,t)$. This is done by numerical
integration of the TDSE using the split-operator-technique~\cite{spo},
with time step of $2.4\times10^{-3}$~fs (or
$0.1$~a.u.). Afterwards, according to the procedure discussed in the previous section (Eqs.~(\ref{eq:modchi})--(\ref{eq:nphase})), we uniquely determine the electronic and
nuclear wave functions in the vector-potential-free gauge.

As an example, we show the evolution of the populations of the BO states in Fig.~\ref{fig: BO} (right panel).

\section{Classical vs. quantum dynamics}\label{sec: cl vs. qm}
We generate classical trajectories by solving Hamilton's equations in the gauge $A(R,t)=0$:
\begin{equation}\label{eqn: hamilton eom}
 \left\lbrace
 \begin{array}{ccl}
  \dot R(t) &=& \dfrac{P(t)}{M} \\
  && \\
  \dot P(t) &=& -\nabla_R E(R,t),
 \end{array}\right.
\end{equation}
using the velocity-Verlet algorithm with the same time step as in the
quantum propagation ($\delta t = 2.4\times10^{-3}$~fs).  It is worth noting that in an actual algorithm the use of such a small time step in not strictly required. As we will show later on, the main features of the TDPES, the steps, form quite smoothly as functions of time, therefore their appearance can be captured also with larger time steps. Moreover, it is worth stressing here, that while $\epsilon_{GI}(R,t)$ and $\epsilon_{GD}(R,t)$ separately have steps, their sum, $\epsilon(R,t)$, is very smooth and capturing its shape within a (fully approximate) numerical scheme will not represent an issue.

In Eq.~(\ref{eqn: hamilton eom}), the energy $E(R)$ is chosen either to be the full TDPES, $\epsilon(R,t)=\epsilon_{GI}(R,t)+\epsilon_{GD}(R,t)$, or the GI part only of the TDPES, $\epsilon_{GI}(R,t)$. We will compare the effect of the resulting dynamics from each. The reason behind this comparison is the feature of the GI part of the TDPES observed in Ref.~\cite{steps}: $\epsilon_{GI}(R,t)$ contains steps that connect its different pieces that are on top of different BOPESs in different slices of R space. This feature from a quasi-classical point of view is reminiscent of the jumps that classical trajectories undergo in TSH. In both cases, the nuclear force is calculated according to different BOPESs depending on the position of the classical trajectory. Therefore, the question may arise as whether $\epsilon_{GI}(R,t)$ alone contains enough information to split the trajectories in two branches, as TSH would do. Moreover, the shape of the GI part of the potential is not affected by the GD part outside the step region, thus the force in the asymptotic regions is given by the slope of the GI part. Thus, one may ask how the energy step in $\epsilon_{GD}(R,t)$ affects the trajectories. Such questions are addressed by evolving classical trajectories on both $\epsilon_{GI}(R,t)$ and $\epsilon(R,t)$.

A set of 2000 trajectories is propagated according to Eqs.~(\ref{eqn:
  hamilton eom}), where the initial conditions are sampled from the
Wigner phase-space distribution corresponding to
$|\chi(R,t=0)|^2=e^{-(R-R_c)^2/\sigma^2}/\sqrt{\pi\sigma^2}$. The
nuclear density at later times is reconstructed from the distribution
of the classical positions and the good agreement with quantum
calculations (as shown below) confirms that the number of
trajectories is sufficient to extract reliable
approximate results.

\begin{figure}[h!]
 \begin{center}
 \includegraphics[angle=270,width=.85\textwidth]{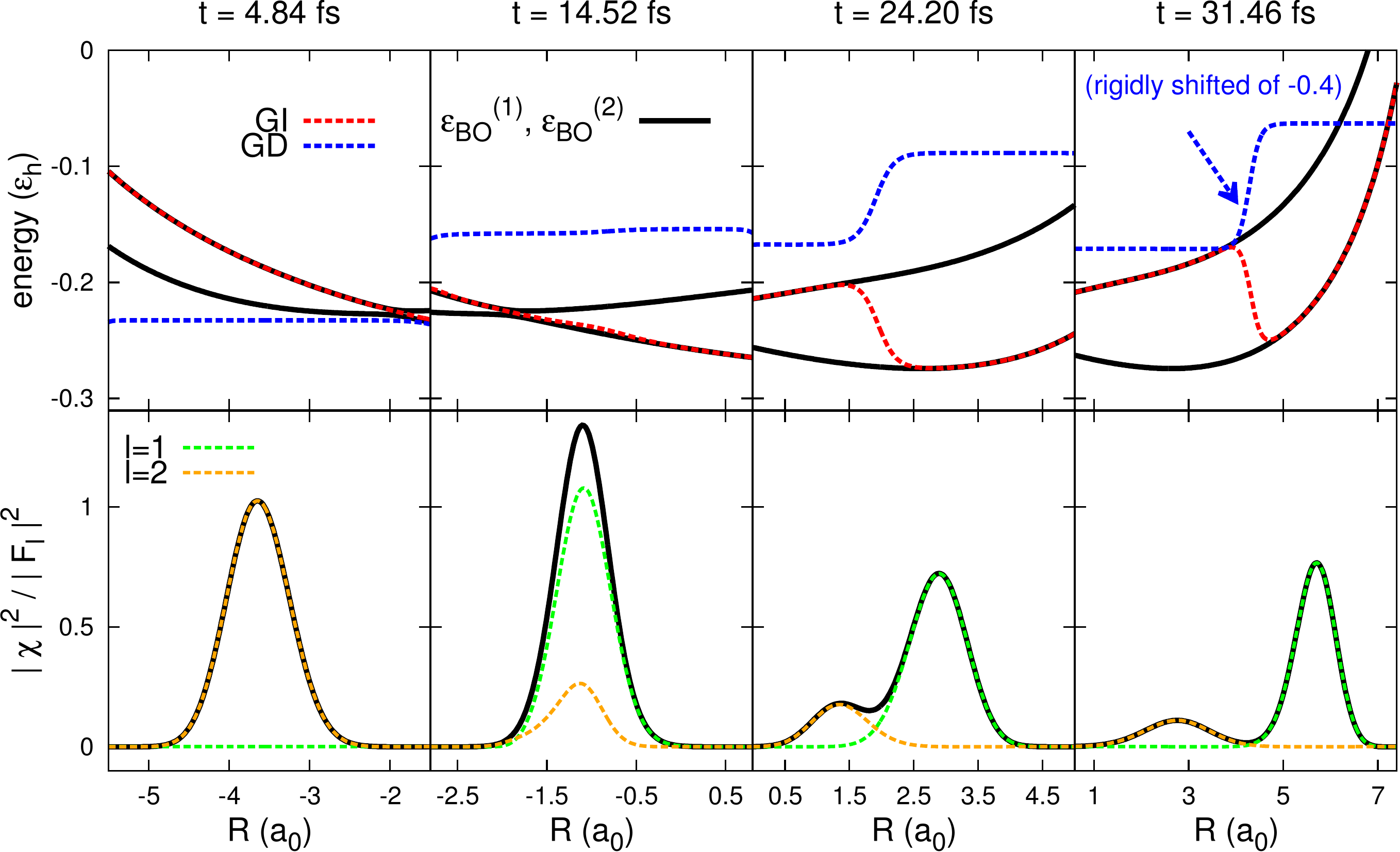}
 \caption{Upper panels: the GI part of the TDPES (cyan) and the GD part (dark-red) (uniformly shifted by -0.4 in all plots), at four times $t=4.84, 14.52, 24.20, 31.46$~fs. The two lowest BO surfaces are shown for reference as solid black lines. Lower panels: the nuclear density $|\chi(R,t)|^2$ (solid black line), and BO-projected densities $|F_1(R,t)|^2$ (dashed green line), $|F_2(R,t)|^2$  (dashed orange). }
 \label{fig: tdpes strong}
 \end{center}
\end{figure}

First we qualitatively analyze the GI and GD components of the TDPES, in connection to the features of the nuclear density. Fig.~\ref{fig: tdpes strong} shows some snapshots of the potentials, $\epsilon_{GI}(R,t)$ and $\epsilon_{GD}(R,t)$ (upper panels), and of the nuclear density, $|\chi(R,t)|^2$, along with its BO-projected components, $|F_1(R,t)|^2$ and $|F_2(R,t)|^2$, (lower panels). The times, as indicated in the figure, are $t=4.84, 14.52, 24.20, 31.46$~fs.

At times $t=4.84, 14.52$~fs, the nuclear wave packet evolves on the
exact potential that either is equivalent to the second adiabatic
surface, on the left of the avoided crossing, or has a diabatic-like
behavior, smoothly connecting the two BO surfaces through the avoided
crossing. At these times, $\epsilon_{GD}(R,t)$ is a constant
function of $R$ and it only produces a physically irrelevant rigid
shift of $\epsilon_{GI}(R,t)$ that has no effect on the dynamics and
could be set to zero. After the nuclear density branches on to the two
surfaces at the avoided crossing (times $t=24.20, 31.46$~fs of
Fig.~\ref{fig: tdpes strong}), both terms of the TDPES develop steps.
The GI surface $\epsilon_{GI}$ lies on one BO surface or the other, with  steps connecting smoothly between the two.
The GD part $\epsilon_{GD}$
is piecewise constant in $R$ space, affecting the dynamics only in the intermediate region where the step joins the two pieces. 
The steps in $\epsilon_{GI}$ and
$\epsilon_{GD}$ appear in the same region and seem to have a similar
slope but with opposite sign.  Still,  the step in $\epsilon_{GD}(R,t)$ does not
exactly cancel the step in $\epsilon_{GI}(R,t)$ (see Section~\ref{sec:
  gd}) and the resulting full TDPES presents a small ``bump'' in this region. It was shown in Refs.~\cite{steps,long_steps} that the steps in the GI and GD parts appear at the
cross-over point, $R_0$, between $|F_1(R,t)|^2$ and $|F_2(R,t)|^2$. 
We will use this symbol $R_0$ from now on to indicate the center of the step region. 
The two branches of the nuclear wave packet undergo different dynamics
because of the different slopes of the (GI part of) the TDPES under each branch,
one being parallel to one BOPES and the other parallel to the other,
while the extent of the splitting critically depends on the combined
effect of the two steps in the GI and GD parts of the exact TDPES, as
we will shortly show.

Knowing the TDPES allows us to directly test the accuracy of a
classical treatment of the nuclei by reconstructing the nuclear dynamics
using classical trajectories.  By evolving an ensemble of multiple
trajectories on the exact TDPES, with a distribution taken from the
exact initial nuclear wave packet (see discussion below Eq.~(\ref{eqn:
  hamilton eom})), we  take into account the quantum uncertainty principle in the initial conditions, so 
the effect of the classical approximation for the
dynamics can be tested by itself, and independently of the approximation used
for the nuclear forces. Further, we can study the impact
of the step structure itself by comparing classical dynamics evolving
on $\epsilon_{GI}(R,t)$ and on $\epsilon(R,t)$.  The difference
between these two potentials, i.e. $\epsilon_{GD}(R,t)$, is piecewise
constant in $R$. Therefore, its shape does not alter the force
calculated from $\epsilon_{GI}(R,t)$ in the regions away from the
step, where $\epsilon_{GI}(R,t)$ is equal to one or the other BO surface.

\begin{figure}[h!]
 \begin{center}
 \includegraphics[angle=270,width=.85\textwidth]{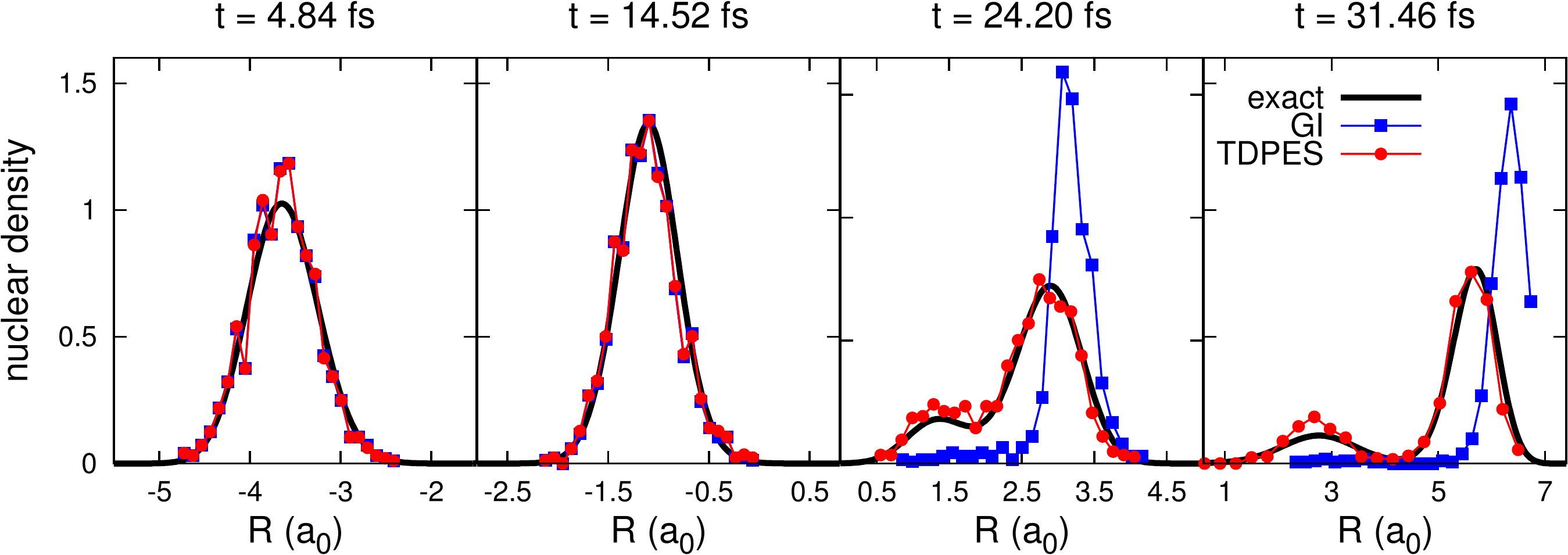}
 \caption{Nuclear density reconstructed from the distribution of the classical positions at four times $t=4.84, 14.52, 24.20, 31.46$~fs, as in  Fig.~\ref{fig: tdpes strong}. The red curves represent the density obtained by propagating classical trajectories on the full TDPES and the blue curves 
 are the results of the propagation on the GI part of the TDPES. For 
 reference, the exact nuclear densities are plotted as black lines.}
 \label{fig: nuclear densities strong}
 \end{center}
\end{figure}

In Fig.~\ref{fig: nuclear densities strong} the nuclear densities 
are approximated with histograms constructed from the distributions of 
classical positions evolving on $\epsilon_{GI}(R,t)$ and on the full TDPES. 
The histograms are represented as blue and red linepoints, respectively, and 
are compared to the exact nuclear density (solid black lines).
Evolving on the GI part of the TDPES yields some effect of splitting,
but the distribution obtained from the propagation of classical
trajectories presents an intense peak localized in the region occupied
by the wave packet propagating on the lower surface, with very little
density on the left of $R_0$. This is clearly not correct, and can be explained
from considering the effect of the step in $\epsilon_{GI}(R,t)$: the
classical trajectories ``slide down'' the step that ``opens'' towards
the lowest BO surface (see snapshots at times $t=24.20$~fs and
$t=31.46$~fs in Fig.~\ref{fig: tdpes strong}), so leaking the
density away from the left to the right.  The propagation of classical
trajectories on the full TDPES confirms the importance of the GD
component of the exact potential, as already suggested in
Ref.~\cite{long_steps}. In Fig.~\ref{fig: nuclear densities strong},
the results from the classical propagation on the exact full TDPES are
in excellent agreement with exact fully quantum results, both before
and after the splitting of the nuclear wave packet. This confirms that 
the splitting of the nuclear wave packet can be captured perfectly with 
treating the nuclei as classical particles, provided the force on the nuclei is the right one.
The steps in both components of the TDPES, yielding a surface with a resultant ``bump'', should be correctly reproduced for an accurate
description.

The step features in the exact TDPES have been observed in previous work~\cite{steps, long_steps} but what is new here is that 
quasiclassical
evolution on the exact TDPES correctly captures the splitting of the
nuclear wave packet and that the balance between the steps in both the GI and GD part is critical to capture the correct dynamics. 
 Our results are based on the fact that we know,
for the simple model of non-adiabatic charge transfer discussed here,
the exact potential that governs the nuclear dynamics. Therefore, we
are able to compute the ``exact'' classical force. It is crucial for
approximations to be able to account for the steps in both $\epsilon_{GI}$
and $\epsilon_{GD}$, that we proved to be
responsible for the correct splitting of the nuclear wave packet and
ensuing dynamics.

We stress once again that the procedure presented so far is not analogous to Bohmian dynamics, as discussed in Appendix~\ref{app: bohmian traj}. When Bohmian trajectories are used to mimic the quantum mechanical evolution of a wave function, an extra contribution, the so-called quantum potential, appears as a time-dependent correction to the standard interaction term, e.g. the Coulomb interaction, in the Hamiltonian. The results shown in Fig.~\ref{fig: nuclear densities strong} are, instead, obtained by only employing the TDPES, without including extra correction terms. The TDPES is the bare potential in the quantum Hamiltonian~(\ref{eqn: nuclear hamiltonian}) and the TDPES is used to calculate classical forces. No additional terms are considered. Therefore, it is not a priori obvious that the classical treatment of nuclear motion is a good approximation even if the TDPES is exact.

\subsection{Nuclear position, momentum and kinetic energy}\label{sec: observables}
Here we  calculate some nuclear observables to demonstrate that measurable quantities can be predicted by evolving classical trajectories under the correct force, i.e. that provided by the exact TDPES. The observable will be evaluated employing the standard expression 
\begin{equation}\label{eqn: average of observables}
O(t) = \frac{1}{N_{traj}}\sum_{I=1}^{N_{traj}} O_I(t) 
\end{equation}
when multiple trajectories (MTs) represent the evolution of the nuclear density. The total number of trajectories is $N_{traj}$, $O_I(t)$ is the value of the observable along the $I$-th trajectory at time $t$ and $O(t)$ is the instantaneous average at time $t$. We will compare Eq.~(\ref{eqn: average of observables}) with the value of the same observable calculated along a single trajectory (ST), whose initial position and momentum are the mean position and momentum of the quantum wave packet and whose evolution is generated according to the force from the TDPES.  As already discussed in Ref.~\cite{long_steps}, a ST cannot capture the spatial splitting of the nuclear density, but can still give information about the observables.

\begin{figure}[h!]
 \begin{center}
 \includegraphics[width=.85\textwidth]{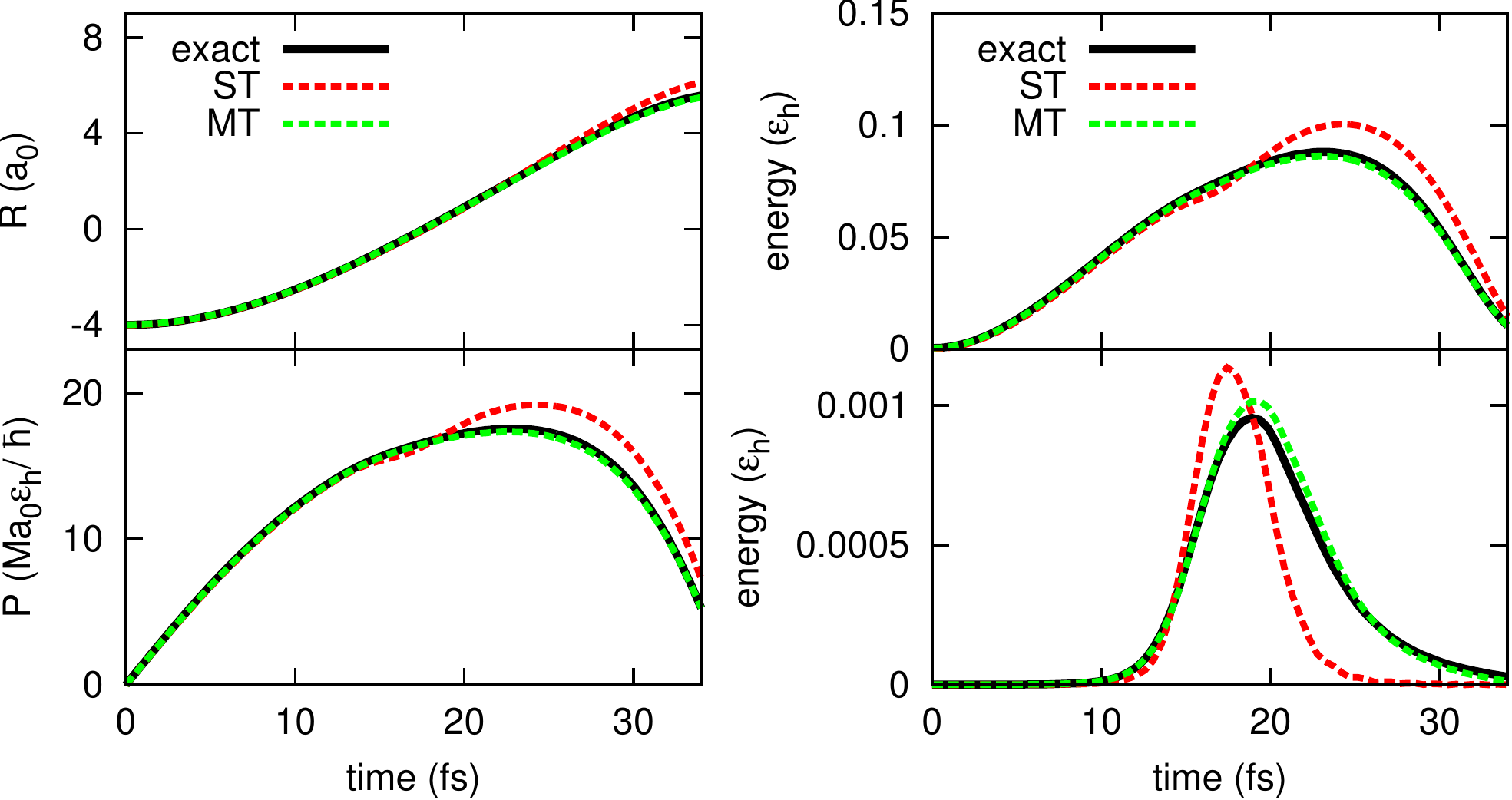}
 \caption{Mean nuclear position (left, upper panel) and mean nuclear momentum 
 (left, lower panel) calculated from the propagation of ST (dashed red line) 
 or of MTs (dashed green line) on the TPDES. 
 Mean nuclear kinetic energy (right, upper panel) and kinetic energy 
 contribution from the term 
 $\hbar^2\left\langle\partial_R\Phi_R(t)\right|\left.\partial_R\Phi_R(t)\right\rangle_r/(2M)$
  (right, lower panel). The black curves represent the quantum mechanical values of the observables.}
 \label{fig: observables}
 \end{center}
\end{figure}

Fig.~\ref{fig: observables} (left panels) shows the mean values of the nuclear position (upper panel) and momentum (lower panel) calculated according to Eq.~(\ref{eqn: average of observables}) for the MT scheme (dashed green lines), compared to the ST (dotted red lines) and exact calculations (continuous black lines). The ST results are not bad, but MT enables an almost perfect agreement with the exact for both the position and momentum variables. This is not surprising given the results presented above. The spatial extension of the nuclear density is accounted for, in an approximated way, in the MT approach and corrects (or washes out) the deviations from ST calculations.

It is instructive to also show the comparison between the nuclear kinetic energy calculated from the full wave function and the corresponding observable estimated from the approximate dynamics. First, we note that, as discussed in Ref.~\cite{AMG2}, when the factorization~(\ref{eqn: factorization}) is used in the expression for the expectation value $T_n(t)$ of the nuclear kinetic energy operator at time $t$
\begin{align}
T_n(t) &= \int d\dulr d\dulR\,\Psi^*(\dulr,\dulR,t)\sum_{\nu=1}^{N_n}\frac{-\hbar^2\nabla_\nu^2}{2M_\nu}\Psi(\dulr,\dulR,t),
\end{align}
the following expression is obtained
\begin{align}
T_n(t) =& \int d\dulR\,\chi^*(\dulR,t) \sum_{\nu=1}^{N_n}\frac{[-i\hbar\nabla_\nu+\bA_\nu(\dulR,t)]^2}{2M_\nu} \chi(\dulR,t) \nonumber\\
-&\int d\dulR\, \left|\chi(\dulR,t)\right|^2\sum_{\nu=1}^{N_n}\frac{\hbar^2}{2M_\nu}\left\langle\nabla_\nu\Phi_\dulR(t)\right|\left.\nabla_\nu\Phi_\dulR(t)\right\rangle_\dulr\nonumber \\
-&\int d\dulR\, \left|\chi(\dulR,t)\right|^2\sum_{\nu=1}^{N_n}  \frac{\bA_\nu^2(\dulR,t)}{2M_\nu}.
\end{align}
In the gauge where $A(R,t)=0$ only two terms survive, namely
\begin{align}\label{eqn: quantum kinetic energy A=0}
T_n(t)=&\frac{-\hbar^2}{2M} \int dR\Big[\chi^*(R,t)\partial_R^2\chi(R,t)\\
&+\left|\chi(R,t)\right|^2\left\langle\partial_R\Phi_R(t)\right|\left.\partial_R\Phi_R(t)\right\rangle_r\Big].\nonumber
\end{align}
In our quasiclassical simulation, this expression is estimated by considering
\begin{align}\label{eqn: approx kinetic energy}
O_I(t)=\frac{P_I^2(t)}{2M}+\frac{\hbar^2}{2M}\left\langle\partial_R\Phi_R(t)\right|\left.\partial_R\Phi_R(t)\right\rangle_r\Big|_{R_I(t)}\nonumber
\end{align}
in Eq.~(\ref{eqn: average of observables}), with the first term the ``bare'' nuclear kinetic energy associated to the $I$-th trajectory at time $t$ and the second term the value of the function $\left\langle\partial_R\Phi_R(t)\right|\left.\partial_R\Phi_R(t)\right\rangle_r$ evaluated at the classical position $R_I(t)$ along the $I$-th trajectory at time $t$. 
Fig.~\ref{fig: observables} (right, upper panel) shows once again that the results from the propagation of MTs on the TDPES perfectly reproduce the quantum expectation value, whereas large deviations are observed for the ST calculation.  The kinetic energy contribution due to the term
$\hbar^2\left\langle\partial_R\Phi_R(t)\right|\left.\partial_R\Phi_R(t)\right\rangle_r/(2M)$, plotted in Fig.~\ref{fig: observables} (right, lower panel), is a small 
fraction of the total kinetic energy, and also is well-approximated by the MT result.

\section{Gauge dependent potentials}\label{sec: gd potentials}
An important result of the previous section is that, while yielding a
zero force in most regions of space, the force from the GD part of the TDPES is
essential to include, due to its step feature, in order to obtain
correct dynamics. Refs.~\cite{steps,long_steps} explained why the slope of the
GI part of the TDPES far from the step tracks that of one or the other BO surface, as well as
how the slope in the step region is related to the slope in the
coefficients as they switch from one surface to the other. Moreover, characteristic features of
the formation and structure of the steps in the GI part of the TDPES were identified.
In particular,
the steps appear in the region around the crossing point $R_0$, which
is where the moduli of the two coefficients of the expansions of
$\Psi(r,R,t)$ (and $\Phi_R(r,t)$) on BO states have the same value,
$\vert F_1(R_0,t)\vert^2 = \vert F_2(R_0,t)\vert^2$, and $\vert
C_1(R_0,t)\vert^2 = \vert C_2(R_0,t)\vert^2 = 1/2$. But what about the
 step-feature in $\epsilon_{GD}$? 
The previous section showed the height of the step in the GD part of
the TDPES must be properly accounted for in order to obtain a
quantitative good estimate of the splitting, so it is  important to characterize its structure as well. In the following
subsections, we estimate the energy difference the step-feature in the GD term represents,
and we look at its effect in terms of velocity-adjustments in
comparison to TSH.

\subsection{Gauge-dependent part of the TDPES}\label{sec: gd}
We observed earlier that steps in the GD part appear at the same position as in the GI term, but have the opposite direction. A ``bump'' then results upon adding the GI and GD term to form the full TDPES. The heights of the steps in these two components of the TDPES are similar but do not
quite cancel each other. We show here why the step in $\epsilon_{GD}$ almost compensates the step in $\epsilon_{GI}$.

First, we prove that $\epsilon_{GD}(R,t)$  has a characteristic sigmoid shape resembling an error-function. This feature is not a result obtained for the particular system studied here or for the particular set of parameters in the Hamiltonian~(\ref{eqn: metiu-hamiltonian}), but it is rather general in the absence of an external time-dependent field. 
We begin by writing the gauge condition $A(R,t)=0$, in terms of the two relevant BO states from Eqs.~(\ref{eqn: vector potential}) and~(\ref{eqn: expansion of Phi})
\begin{align}
 0=\sum_{l=1,2}&\left|C_l(R,t)\right|^2\partial_R\gamma_l(R,t)-\frac{i\hbar}{2}\partial_R\sum_{l=1,2}\left|C_l(R,t)\right|^2\nonumber\\
&-i\hbar\sum_{l,k=1,2} C_l^*(R,t)C_k(R,t)d_{lk}(R),
\end{align}
where the symbol $\gamma_l(R,t)$ has been used to indicate the phase of the coefficient $C_l(R,t)$ of the electronic wave function (Eq.~(\ref{eqn: expansion of Phi})) and $d_{lk}(R)=\langle\varphi^{(l)}_R|\partial_R\varphi_R^{(k)}\rangle_r$ indicates the non-adiabatic coupling vectors (NACVs) in 1D. 
The second term on the right-hand-side is identically zero, due to the PNC in Eq.~(\ref{eqn: PNC on BO}), while the third term can be neglected in the region far from the avoided crossing where the NACVs are negligible. This is the region we are interested in, since $\epsilon_{GD}(R,t)$ is different from a constant function only after the passage of the nuclear wave packet through the avoided crossing. The remaining term gives
\begin{equation}
 \left|C_1(R,t)\right|^2\partial_R\gamma_1(R,t) = -\left|C_2(R,t)\right|^2\partial_R\gamma_2(R,t),
\end{equation}
which means $\partial_R\gamma_1(R,t)=0$ $\forall \,R$ where $|C_1(R,t)|^2=1$ while $|C_2(R,t)|^2=0$. Similarly, $\partial_R\gamma_2(R,t)=0$ $\forall \,R$ where $|C_2(R,t)|^2=1$ while $|C_1(R,t)|^2=0$. We conclude $\gamma_l(R,t)=\Gamma_l(t)$, namely the phase of the coefficient $C_l(R,t)$ is only a function of time (constant in space), in the region where the squared modulus of the corresponding coefficient is equal to unity.
\begin{figure}[h!]
 \begin{center}
 \includegraphics[width=.45\textwidth,angle=270]{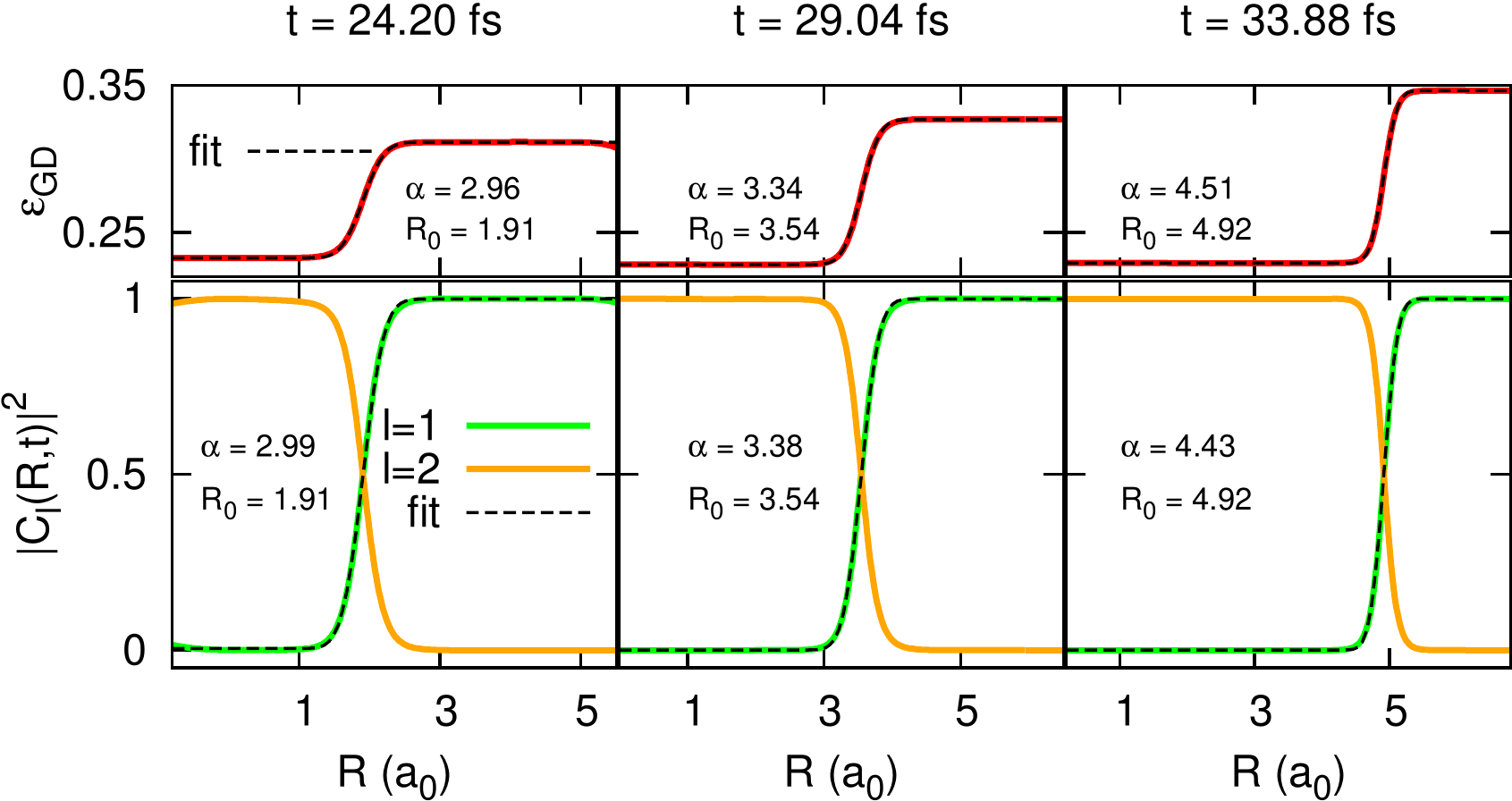}
 \caption{Upper panels: $\epsilon_{GD}(R,t)$ plotted as red line at some time steps after the nuclear wave packet has crossed the non-adiabatic coupling region. The fitting function in Eq.~(\ref{eqn: fitting function}) is shown as black dashed line. Lower panels: $|C_l(R,t)|^2$ for $l=1$ (green lines) and $l=2$ (orange lines) at the same time steps, with the fitting function shown again as dashed black line. The fitting parameters $\alpha(t)$ and $R_0(t)$ from Eq.~(\ref{eqn: fitting function}) are indicated in the upper and lower panels.}
 \label{fig: coefficients and gd}
 \end{center}
\end{figure}
We confirm this by showing in Fig.~\ref{fig: coefficients and gd} (lower panels) some snapshots of the functions $|C_1(R,t)|^2$ and $|C_2(R,t)|^2$. In the regions where one of the two coefficients is equal to 1, the other is 0. This means that the electronic wave function has ``collapsed'' onto one adiabatic state, that whose coefficient is non-zero.  Eq.~(\ref{eqn: expansion of Phi}) can then be written as
\begin{align}
\Phi_R(r,t)=\left\lbrace
\begin{array}{ll}
e^{\frac{i}{\hbar}\Gamma_l(t)}\varphi_R^{(l)}(r) & \forall\,R:\left|C_l(R,t)\right|^2=1 \\
\sum_{l}C_l(R,t)\varphi_R^{(l)}(r)& \mbox{otherwise},
\end{array}\right.
\end{align}
meaning that $\Phi_R(r,t)$ has a purely adiabatic character for $R$ where $|C_l(R,t)|^2$'s are either zero or one, while a linear combination of adiabatic states in the region in between. The sigmoid-shape of $\epsilon_{GD}(R,t)$ then follows by using Eq.~(\ref{eqn: gd tdpes}),
\begin{align}
\epsilon_{GD}(R,t)=\left\lbrace
\begin{array}{ll}
\dot\Gamma_l(t) & \forall\,R:\left|C_l(R,t)\right|^2=1 \\
\sum_{l}|C_l(R,t)|^2\dot\gamma_l(R,t)& \mbox{otherwise}.
\end{array}\right.
\end{align}
Numerical results are shown at different time steps in Fig.~\ref{fig: coefficients and gd} (upper panels).

The ``collapse'' of the electronic wave function on one or the other adiabatic state is a smooth and continuous process that can be observed in time by looking either at the step formation in the GI (and GD) part of the TDPES, for instance in Fig.~\ref{fig: tdpes strong}, or at the step formation in $|C_l(R,t)|^2$ (with $l=1,2$), in Fig.~\ref{fig: coefficients and gd}.

Both $\epsilon_{GD}(R,t)$ and $|C_l(R,t)|^2$ can be fitted via a function $f(R,t)$ of the form
\begin{equation}\label{eqn: fitting function}
f(R,t)= o(t)+h(t) \,\mathrm{erf} \left[\alpha(t)\left(R-R_0(t)\right)\right],
\end{equation}
where the time-dependent parameters $o(t), h(t), \alpha(t)$ and $R_0(t)$ are determined by the fitting procedure.  Fig.~\ref{fig: coefficients and gd} shows as thin dashed black line the analytic function Eq.~(\ref{eqn: fitting function}) and the values of the parameters indicating the slope of the step, $\alpha(t)$, and the position of the step, $R_0(t)$. In the lower panels we need only to fit  $|C_1(R,t)|^2$, since due to the PNC,  $|C_2(R,t)|^2=1-|C_1(R,t)|^2$. We observe that the values of the fitting parameters are very similar for $|C_l(R,t)|^2$ and $\epsilon_{GD}(R,t)$, meaning that the step in  $\epsilon_{GD}(R,t)$ has the same spatial-dependence as that of  $|C_l(R,t)|^2$. In the case of $|C_1(R,t)|^2$ the remaining two parameters, $o(t)=h(t)=0.5$,  as expected. In the case of $\epsilon_{GD}(R,t)$, we analytically derive in the appendix that 
\begin{align}
2h(t)\simeq&\frac{\alpha}{\sqrt{\pi}}\int dR \left[\epsilon_{BO}^{(2)}(R)-\epsilon_{BO}^{(1)}(R)\right]
\,e^{-\alpha^2(t)(R-R_0(t))^2},\label{eqn: leading term bis}
\end{align}
while we do not consider $o(t)$ since it is only a uniform potential, that has no physical effect on the dynamics. Here, the integral is computed with boundaries $\pm \infty$ and we used the property that $|C_1(R,t)|^2$ has an error-function shape in order to analytically determine its spatial derivative as a Gaussian.

\begin{figure}[h!]
 \begin{center}
 \includegraphics[angle=270,width=.6\textwidth]{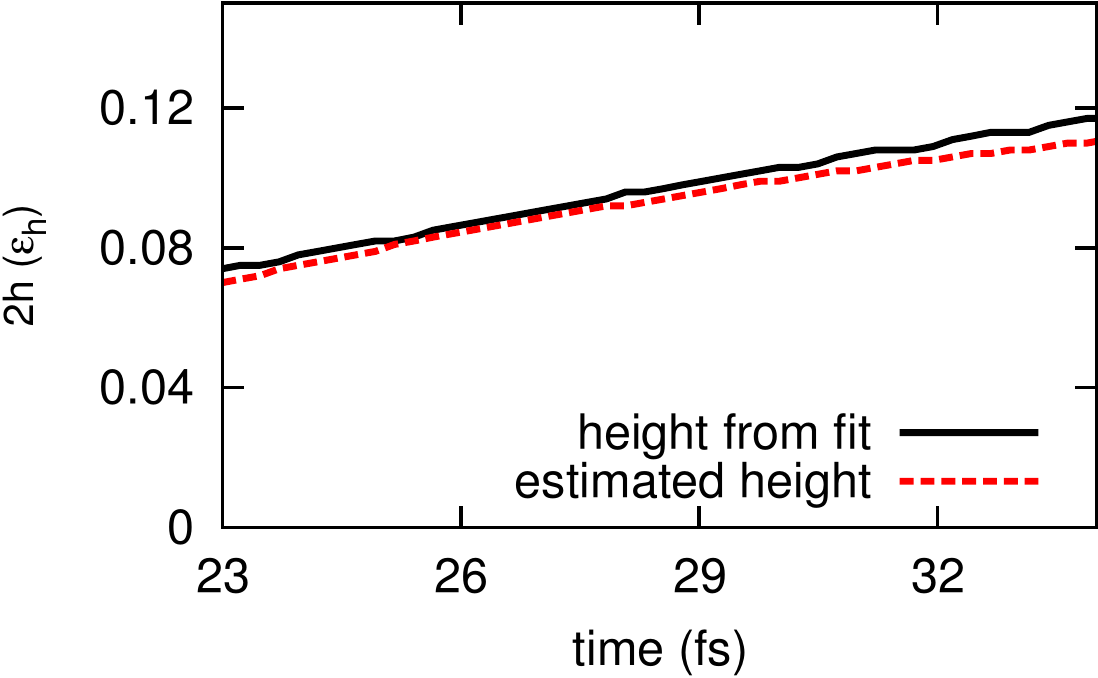}
 \caption{Comparison between the height of the step in $\epsilon_{GD}(R,t)$ estimated from Eq.~(\ref{eqn: leading term bis}) (red) and the value calculated from the fitting (black) with an error-function.}
 \label{fig: gd height}
 \end{center}
\end{figure}

Numerical results are shown in Fig.~\ref{fig: gd height}, where we compare the value of the height obtained from the fit of $\epsilon_{GD}$ via the error-function in Eq.~(\ref{eqn: fitting function}) and from Eq.~(\ref{eqn: leading term bis}).
Fig.~\ref{fig: gd height} shows that the results of our estimate are in good agreement with the fit. 

In summary, we have explained why $\epsilon_{GD}$ is piecewise constant, with a sigmoid structure whose height is an average of the difference between the BO energies weighted by a Gaussian centered in the step region, Eq.~(\ref{eqn: leading term bis}). This explains (i) why the step in $\epsilon_{GD}(R,t)$ almost compensates the energy difference in $\epsilon_{GI}(R,t)$ and (ii) why the slope in the GD part of the TDPES has the opposite sign with respect to the GI part.

\subsection{Vector potential}\label{sec: A}
The analysis in the previous sections shows that the GD component of
the TDPES, $\epsilon_{GD}$, does not affect the force that evolves the classical
trajectories on either side of the step (it is constant in these
regions). However, it diminishes the energy separation
in $\epsilon_{GI}$ between the two sides. This energy barrier almost
disappears in the full TDPES. However, the TDPES is gauge-dependent, and the question arises as to how does  this reduction of the energy difference that we see in $\epsilon_{GD}$ appear in other gauges? In particular, going to a gauge where
$\epsilon_{GD}(R,t)=0$ means that the non-zero vector potential must compensate
the effect of the energy step in the GI part of the TDPES. We will analyze this effect in this section. 

The gauge $\epsilon_{GD}(R,t)=0$ offers an interesting point of view,
giving perhaps a more direct interpretation of the TSH scheme in terms
of the exact TDPES and vector potential.  In TSH, the force that
produces the nuclear evolution is given by the gradient of one of the
BOPESs, and the classical particles evolve adiabatically on the BO
surfaces before and after the stochastic jumps take place.
In our exact formulation in this gauge, a large component of the classical force driving
the nuclear motion is given by the GI part of the exact TDPES, which reduces to  the gradient of either one or the other adiabatic surface away from the step region, similar to  TSH. But there is also a component to the force from the time-dependent vector potential. This appears as a momentum correction, whose effect and interpretation in the perspective of TSH will be shown below.

We apply the gauge-function $\theta(R,t)$ (see Eqs.~(\ref{eqn: gauge}),~(\ref{eqn: transformation of epsilon}) and~(\ref{eqn: transformation of A})) on the wave functions and potentials of the previous sections, such that in the new gauge, 
\begin{equation}
\tilde\epsilon_{GD}(R,t) = \epsilon_{GD}(R,t)+\dot\theta(R,t)=0\,.
\end{equation}
Equivalently,
\begin{equation}
\theta(R,t)=-\int^tdt'\,\epsilon_{GD}(R,t').
\end{equation}
From Eq.~(\ref{eqn: transformation of A}), noting that in the previous gauge $A(R,t) = 0$,
\begin{equation}\label{eqn: A from gd}
\tilde A(R,t) = \int_0^tdt'\,\Big(-\partial_R\epsilon_{GD}(R,t')\Big).
\end{equation}
That is, the vector potential in the gauge where it absorbs all the
gauge-dependence, is the time integral of the force
generated by the GD part of the TDPES in the gauge where the vector potential is zero.

\begin{figure}[h!]
 \begin{center}
 \includegraphics[width=.85\textwidth]{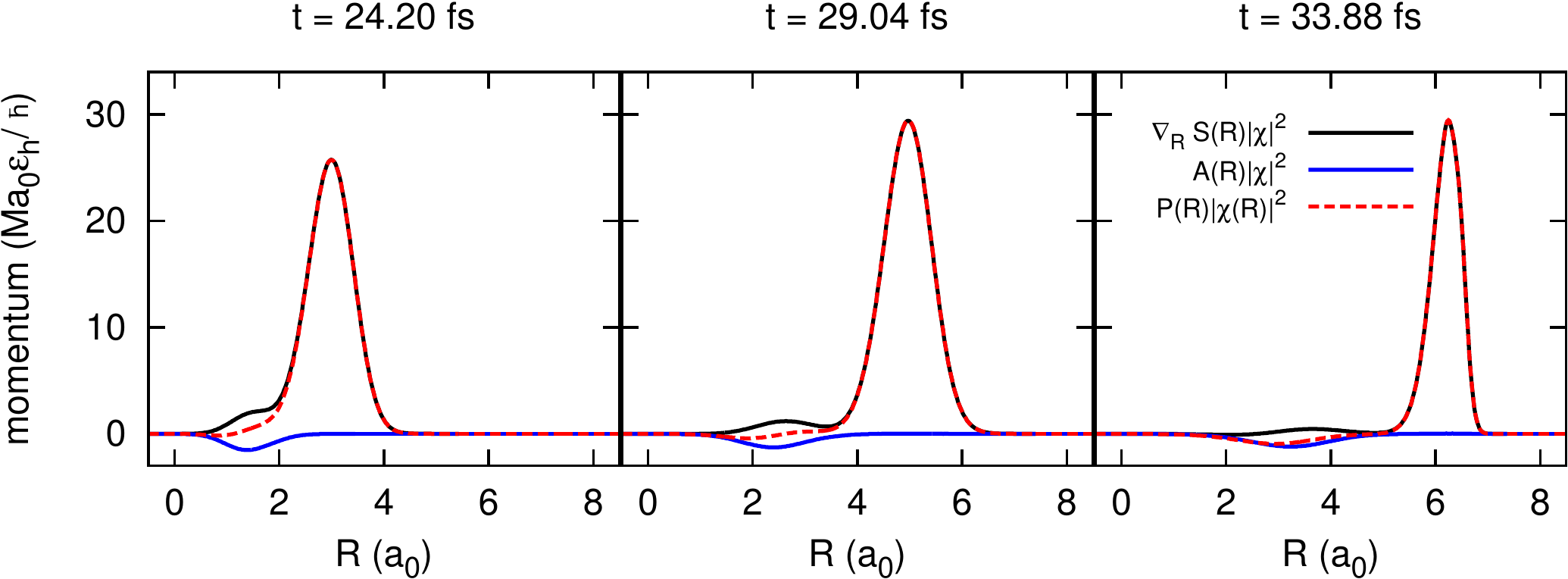}
 \caption{Contributions, at times $t=25.41,30.25,33.88$~fs, to the total nuclear momentum density from the terms $\nabla_R \tilde S(R,t)$ (black line) and $\tilde A(R,t)$ (blue line), from Eq.~(\ref{eqn: P and A}), weighted by the nuclear density. The dashed-red line represents the term on the left-hand-side in Eq.~(\ref{eqn: P and A}), weighted by the nuclear density.}
 \label{fig: momentum corrections}
 \end{center}
\end{figure}

The vector potential contributes a term to the nuclear momentum density, that is induced by the coupling to the electrons: 
\begin{equation}\label{eqn: P and A}
P(R,t)=\nabla_R \tilde{S}(R,t)+\tilde{A}(R,t)
\end{equation}
where $\tilde{S}$ is the phase of the nuclear wave function in the present gauge.
Note that the total averaged momentum is the integral of the total current density: $P(t) = \int dR\, j(R,t) = \int dR\,\vert\chi(R,t)\vert^2 P(R,t)$, plotted in Fig.~\ref{fig: momentum corrections}.
Both contributions, $\nabla_R \tilde S(R)$ and $\tilde A(R)$, weighted by the nuclear density, are shown in Fig.~\ref{fig: momentum corrections} as the black and blue lines. 
Examining alongside $\epsilon_{GI}$ shown in Fig.~\ref{fig: tdpes strong}, we observe that the vector potential lowers the momentum and kinetic energy in the region where the nuclear wave packet evolves on the upper surface.

In TSH different adiabatic surfaces are energetically accessible by
the classical nuclei because of the stochastic jumps and the
subsequent momentum rescaling: when a jump occurs in the direction of
increasing potential energy, e.g. from state 1 to state 2 in our
example, the velocity of the classical particle is reduced by the
amount determined by imposing energy conservation, along the direction
of the NACV between the states involved in the transition. In our scheme  based on the exact TDPES, we see a similar effect in two different ways depending on the gauge: either
the GD part of the potential is responsible for bringing
``energetically closer'' different BOPES, or the vector potential provides
the necessary kinetic energy contribution. From the observations
reported in the present section, the vector potential contributes to
the nuclear momentum as a reduction of the propagation velocity of the
trajectories on the upper surface, with respect to the trajectories on the
lower surface. This is reminiscent of the momentum adjustment in TSH,
when the classical particles undergo a non-adiabatic jump from one
BO surface to the other.  An interesting development of these
qualitative observations would be the analysis, in higher dimensions,
of the direction of the momentum adjustment due to the vector
potential, in comparison to the direction chosen in the TSH approach,
i.e. the direction of the NACVs. This line of investigation,
along with a more quantitative understanding of the connection between
the time-dependent vector potential and the velocity corrections in
the TSH scheme is currently under investigation.

\section{Further relation with TSH}\label{sec: decoherence}
A consequence of the steps in the GI and GD components of the TDPES is that they allow the nuclear wave packet to feel forces from different BOPESs in different regions of space at the same instant in time. This feature signals electronic decoherence. Although TSH shares the feature that wave packets evolving in different spatial regions on different surfaces experience different forces, it does not capture the electronic decoherence that should come hand-in-hand. There has been extensive and on-going developments to build decoherence into TSH~\cite{shenvi-subotnikJCP2011, landryJCP2013_2, prezhdoJCP2012, truhlarFD2004, truhlarACR2006, kapralJCP2008, persicoJCP2007} 
but it remains a challenging problem today. We now ask what we can learn about decoherence from the exact TDPES in relation to TSH. We will find that the concerted action of the step features in $\epsilon_{GI}$ and $\epsilon_{GD}$ results in decoherence. 
 
Consider the force produced by the GD part of the potential. The force is a gauge-independent quantity, therefore does not depend on the choice of the gauge. In a gauge where there is zero vector potential, the classical nuclear force is the sum of two contributions, 
\begin{align}\label{eqn: classical force}
-\partial_R\epsilon(R,t)=-\partial_R\epsilon_{GI}(R,t)-\partial_R\epsilon_{GD}(R,t),
\end{align}
where the second term on the right-hand-side can be also written as
\begin{align}\label{eqn: force from A}
-\partial_R\epsilon_{GD}(R,t)=\partial_t \tilde A(R,t),
\end{align}
 in the gauge where the entire GD term has been transformed to a vector potential, Eq.~(\ref{eqn: A from gd}). The results shown in Fig.~\ref{fig: coefficients and gd} for $\epsilon_{GD}(R,t)$ clearly indicate that Eq.~(\ref{eqn: force from A}) either is zero or has a Gaussian-like shape, since $\epsilon_{GD}(R,t)$ either is constant or has the characteristic sigmoid shape reminiscent of an error-function.

We now consider single trajectories generated according to the fewest-switches algorithm, on one hand, and the trajectories evolving on the TDPES, on the other hand. We distinguish two classes of trajectories: (i) those that in TSH do not hop, thus always propagating on the upper BO surface $\epsilon_{BO}^{(2)}(R)$ and (ii) those that undergo a single hop, whose evolution after the passage through the avoided crossing takes place along the lower surface $\epsilon_{BO}^{(1)}(R)$. In this analysis we do not look at all other trajectories, undergoing two or more hops.
\begin{figure}[h!]
 \begin{center}
 \includegraphics[width=.7\textwidth]{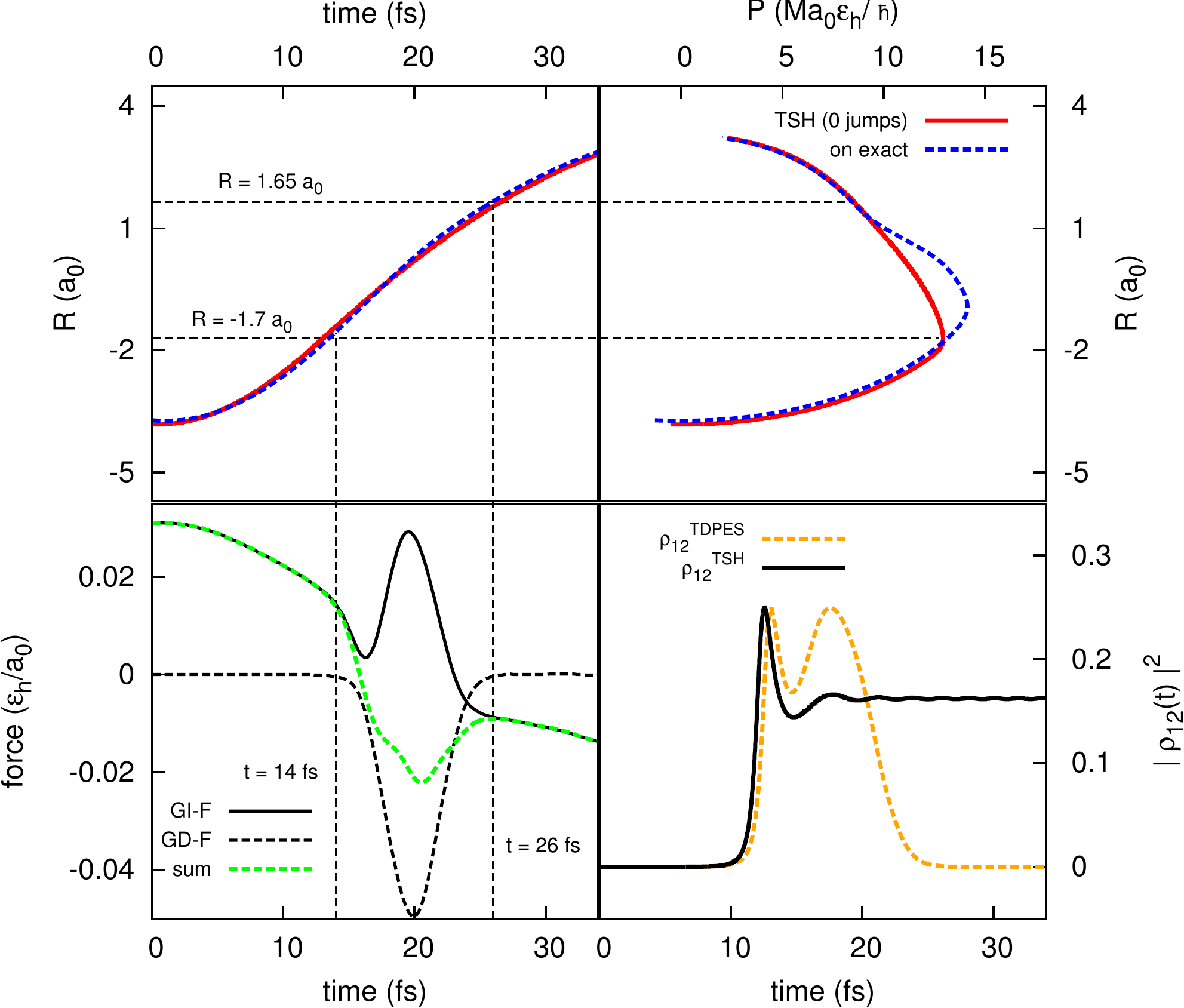}
 \caption{Results for the trajectories of class (i). Upper left panel: trajectory calculated from TSH (red line) and by propagating classical positions on the TDPES (dashed blue line). Upper right panel: phase space (momentum shown on the $x$-axis and position on $y$-axis) from TSH (red line) and from the trajectory on the TDPES (blue line). Lower left panel: the GI-force (continuous black line), the GD-force (dashed black line) and their sum (dashed green line), determined by evaluating Eqs.~(\ref{eqn: classical force}) and~(\ref{eqn: force from A}) at the instantaneous position. Lower right panel: squared moduli of the off-diagonal elements of the electronic density matrix $\rho_{12}^{\mathrm{TDPES}}(t)=|C_1(R_{cl}(t),t)|^2|C_2(R_{cl}(t),t)|^2$, evaluated at the classical positions (dashed orange lines), compared to the corresponding quantity $\rho_{12}^{\mathrm{TSH}}(t)$ (black lines) calculated along the TSH trajectory. The vertical and horizontal dashed black lines highlight the regions of time and space, respectively, where the effect of the vector potential on the classical trajectory is not zero. The values associated to these regions are also reported in the plots.}
 \label{fig: strong upper}
 \end{center}
\end{figure}

We have identified a typical trajectory from each of these classes and matched each up with a classical trajectory evolving on the exact TDPES, that has almost identical initial momentum and position and that either ``ends up'' on the upper surface (class (i)) or slides over to the lower BO surface (class (ii)). Since TSH is a stochastic scheme, identical trajectories are difficult to find. Nevertheless, this comparison allows us to classify also the trajectories on the TDPES as class (i) or (ii).

\begin{figure}[h!]
 \begin{center}
 \includegraphics[width=.7\textwidth]{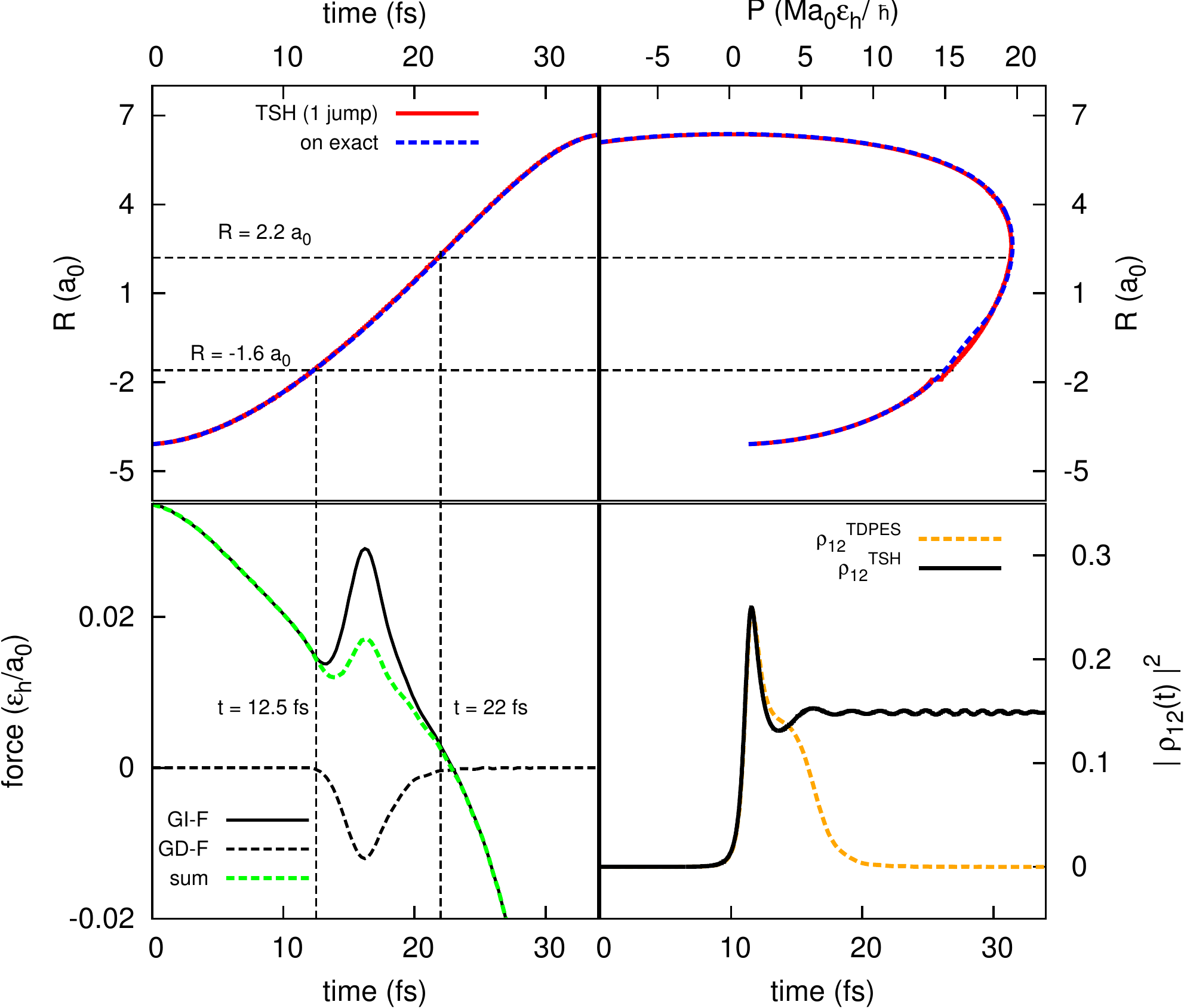}
 \caption{Same as in Fig.~\ref{fig: strong upper} but for the trajectory of class (ii).}
 \label{fig: strong lower}
 \end{center}
\end{figure}

These pairs of trajectories are plotted  in the upper left panel of Figs.~\ref{fig: strong upper} (class (i)) and~\ref{fig: strong lower} (class (ii)). These panels verify the similarity of the TSH (red lines) and TDPES (dashed blue lines) trajectories of each pair. The other panels represent: (upper right) the phase space from TSH and the TDPES (in this case the momentum is shown on the $x$-axis and the position on the $y$-axis); (lower left) the force from the first term on the right-hand-side of Eq.~(\ref{eqn: classical force}), in the following referred to as GI-force, (continuous black line), the force in Eq.~(\ref{eqn: force from A}), referred to as GD-force, and their sum (dashed green line), determined by evaluating Eqs.~(\ref{eqn: classical force}) and~(\ref{eqn: force from A}) at each time at the classical position along the trajectory propagating on the TDPES; (lower right) the squared moduli of the off-diagonal elements of the electronic density matrix $\rho_{12}^{\mathrm{TDPES}}(t)=|C_1(R_{cl}(t),t)|^2|C_2(R_{cl}(t),t)|^2$, evaluated at the classical positions (dashed orange lines), as the force in the lower left panel, compared to the corresponding quantity $\rho_{12}^{\mathrm{TSH}}(t)$ (black lines) calculated along the single TSH trajectory shown in the upper-left panels.

With the classification of the trajectories in mind, we compare the phase space of the trajectories associated to the upper and to the lower surfaces. Clearly, for a trajectory in class (ii), the momentum calculated according to TSH will show a discontinuous behavior in the vicinity of the avoided crossing, because it undergoes a hop from $\epsilon_{BO}^{(2)}(R)$ to $\epsilon_{BO}^{(1)}(R)$, while the trajectory propagating on the TDPES continues to evolve smoothly. This is indeed verified in the upper right panel of Fig.~\ref{fig: strong lower} where the ``kink'' in the red curve near $R=-2$~$a_0$ indicates a momentum jump occuring at the corresponding surface hop. However, after passing through this region, the curves are once again very close to each other.

This suggests that the {\it time-integrated} effects of the forces from the  TSH surface-hop and the exact vector potential on the classical trajectory are  approximately the same, although their nature is rather different. Both the TSH trajectory and the TDPES trajectory begin evolving under the force from the upper BO surface, but, due to  the velocity adjustment, the  TSH trajectory experiences an instantaneous $\delta$-function-like force when it hops between the  surfaces.  We can interpret this force as arising from a vector potential that turns on sharply  at the avoided crossing. The force on the  TDPES trajectory, on the other hand, is always smooth. In the lower left panels of Figs.~\ref{fig: strong upper} and~\ref{fig: strong lower} we observe a peculiar behavior of the classical force in the region within the two dashed vertical lines: outside this region, the GD-force is zero, suggesting that classical dynamics is completely governed by one or the other BOPES; within the region, the effect of the steps in $\epsilon_{GI}$ and $\epsilon_{GD}$ (or the appearance of the vector potential according to Eq.~(\ref{eqn: force from A})) is clearly indicated by the peaks. The time intervals indicated by the vertical dashed black lines in Figs.~\ref{fig: strong upper} and~\ref{fig: strong lower} give an estimate of the time during which the GD-force has a non-zero effect on the classical trajectory. The forces in TSH coincide with those of the TDPES only outside these time intervals. 
One can see that such intervals coincide with space regions where the phase space from TSH calculations and resulting from the trajectories evolving on the TDPES are qualitatively different. These regions are indicated by the horizontal dashed black lines.

This time window during the gradual onset and offset of the GD-force infinitely reduces when interpreted in the light of TSH. Although this difference has little net effect on the classical dynamics after the transition region (the two trajectories end up matching up again), it has a profound influence on the electronic amplitudes associated  with these trajectories, as we will now argue.  The major consequence of the appearance of the peaks in the GI-force and the GD-force on the TDPES-evolved trajectory, and the associated structure in $\hat U_{en}^{coup}$ in the electronic equation, is to induce decoherence, an effect that the instantaneous action of the force on the TSH-evolved trajectory does not capture properly. From a quantum mechanical point of view, we expect to observe decoherence when the components of the nuclear wave packet associated to different BO states evolve independently from each other on the two adiabatic surfaces. In the mixed quantum-classical picture provided by the classical trajectories propagating on the TDPES, decoherence is correctly observed. In order to quantify this statement, we employ, as indicator of decoherence, the off-diagonal elements of the electronic density matrix, i.e. $|\rho_{12}^{\mathrm{TDPES}}(t)|^2=|C_1(R_{cl}(t),t)|^2|C_2(R_{cl}(t),t)|^2$ (dashed orange lines) in the lower right panels of Figs.~\ref{fig: strong upper} and~\ref{fig: strong lower}. The coefficients $C_j(R,t)$ with $j=1,2$, from the expansion of the electronic wave function on the adiabatic basis, are functions of position and time, and in the expression of $|\rho_{12}^{\mathrm{TDPES}}(t)|^2$ they are evaluated at each time at the classical positions along the trajectory $R_{cl}(t)$. The elements of the density matrix $|\rho_{12}^{\mathrm{TDPES}}(t)|^2$ are compared with the corresponding quantity calculated along a given TSH trajectory (black lines in the lower right panels of Figs.~\ref{fig: strong upper}  and~\ref{fig: strong lower}), i.e. $|\rho_{12}^{\mathrm{TSH}}(t)|^2$.

In the same time intervals indicated by the vertical lines in the lower left panels, $|\rho_{12}^{\mathrm{TDPES}}(t)|^2$ decays to 0. These time intervals, and the corresponding regions of space, can be associated to the decay of the coherences ($\rho_{12}$). Now if we compare these matrix elements with those calculated along a given trajectory according to the TSH scheme, the decay to 0 is clearly absent. After the transition through the coupling region, TSH predicts for all trajectories approximately the same values of $|\rho_{12}^{\mathrm{TSH}}(t)|^2$, which, being different from zero, imply a spurious coherence.

Since the velocity adjustment in TSH is analogous to the action of an instantaneous vector potential (or an infinitely steep step in the TDPES), decoherence is not induced after the nuclear wave packet, i.e. the trajectory-bundle, crosses the coupling region. The GD-force and its combined effect with the GI-force can be used to estimate a decoherence time and length and to provide corrections to the problems of TSH. Further analysis in this direction will be investigated.

\section{Conclusions}\label{sec: conclusions}
This work shows  the \textit{best} results that can be obtained when an (independent) trajectory-based mixed quantum-classical procedure is used to simulate the coupled non-adiabatic dynamics of electrons and nuclei. Here, the electronic effect on the classical nuclei is not approximated, but it is treated exactly by employing the TDPES and the time-dependent vector potential defined in the framework of the exact factorization of the electron-nuclear wave function of Refs.~\cite{AMG,AMG2}. A quasiclassical treatment of nuclear dynamics is sufficient to accurately capture the true quantum dynamics, in particular the splitting of the nuclear wave packet. Critical step and bump features of the exact TDPES are responsible for the correct dynamics, and these were analysed in detail.

Envisaging the development of a mixed quantum-classical
algorithm from the analysis reported here, we observe that reproducing
the step feature in the GI and GD components of the TDPES is crucial
to correctly reproduce the splitting of a nuclear wave packet created by an
avoided crossing. In particular, after passage through the avoided crossing, the correct dynamics of the different branches of the nuclear wave packet
are caused by the
deformation of the full time-dependent potential, that (i) becomes
parallel to one or the other BO surface in different regions and (ii)
develops a small bump in the intermediate region. 
A detailed analysis of the shape of the GD component was given here, while earlier work had focussed on the GI component~\cite{steps,long_steps}. In particular, we explained why the shape is a sigmoid, and why the step in the GD component largely cancels that in the GI component. 
The effect of the
GD part of the TDPES was analyzed also under a different
perspective, by pointing out the qualitative connection with the
\textit{momentum adjustment} in the TSH procedure. With a suitable
change of the gauge, i.e. $\epsilon_{GD}(R,t)=0$, the time-dependent
vector potential appears as a kinetic contribution in the nuclear
Hamiltonian and its effect is reminiscent of the velocity rescaling of
TSH.

We have also discussed the role of the steps in the two components of the TDPES for the correct account of decoherence. The steps indicate that both the GI-force and GD-force develop a peak, giving a time-interval corresponding to the typical time for the decay of coherences. These peaks, or equally the steps in the potentials, are essential to reproduce electronic decoherence.
 
Future work will further highlight differences and analogies to existing procedures. For instance, the decoherence force proposed in the decay-of-mixing method by Truhlar and co-workers~\cite{truhlarFD2004, truhlarACR2006} 
seems to play a similar role as the force (the peaks in the GI-force and in the GD-force) arising during the decoherence time window. Also, like dynamics on the exact TDPES, the non-adiabatic dynamics scheme based on the Meyer-Miller-Stock-Thoss (MMST) Hamiltonian~\cite{millerJCP1997, thossPRL1997} also has classical trajectories evolving on just one potential, and it was shown that when these Ehrenfest trajectories are treated semiclassically, including phase information, then wave packet splitting can occur~\cite{millerJCP2007, millerJPCA2009, millerJPCA2013}. Comparisons of mixed quantum-classical schemes based on the exact factorization and methods such as these will lead to more insight into electron-nuclear correlation. 

At longer times, not studied here, when multiple passes through
avoided crossing regions and reflections become important, the
quasiclassical approach is expected to become poor. However, what our
work here demonstrates is that it certainly is able to fundamentally capture the
 non-adiabatic charge transfer event. A modified
approach, based again on an ensemble of classical trajectories but now
incorporating phases, such as in the semiclassical dynamics approaches of
Refs.~\cite{millerJPCA2013, heller}, could prove to be a promising approach in such
cases, and is an avenue for  further research. 

The observations that we have reported here establish the basis for
interpreting existing approximated methods that deal with the problem
of coupled electron-nuclear dynamics and for understanding how their
deficiencies can be cured. Furthermore, our work leads to new insights
into the physics of electronic non-adiabatic processes and may lead
to the design of new mixed quantum-classical algorithms~\cite{mqc, long_mqc} that satisfy
exact requirements.

\section*{Acknowledgements}
Partial support from the Deutsche Forschungsgemeinschaft (SFB 762) and from the European Commission (FP7-NMP-CRONOS) and the US Department of Energy Office of Basic Energy Sciences, Division of Chemical Sciences, Geosciences and Biosciences under Award DE- SC0008623 is gratefully acknowledged.

\appendix

\section{Connection to Bohmian trajectories}\label{app: bohmian traj}
We devote this appendix to a brief overview of the Bohmian approach~\cite{wyattJCP2001, wyattJCP2002} to quantum dynamics. Even though this procedure might seem somehow connected to the work presented in the paper, as it is based on the propagation of trajectories on a time-dependent potential, it will become clear here that this is not the case. Also, let us discuss only the adiabatic case, since the non-adiabatic version of Bohmian dynamics, recently proposed by Tavernelli and co-workers~\cite{tavernelliPCCP2011}, is based on the Born-Huang expansion of $\Psi(\dulr,\dulR,t)$ and it is used the evolve the coefficients of this expansion rather than the wave function.

The state of the system is described by the wave function $\chi(\dulR,t)$, that evolves according to the TDSE
\begin{equation}\label{eqn: tdse for chi}
\hat H \chi(\dulR,t) = i\hbar \partial_t\chi(\dulR,t),
\end{equation}
with Hamiltonian $\hat H$ containing a kinetic energy term and a potential term $\hat V(\dulR)$, e.g. the (static) Coulomb interaction among particles. For the moment, no relation between $\chi(\dulR,t)$ in Eq.~(\ref{eqn: tdse for chi}) and the nuclear wave function of Eq.~(\ref{eqn: factorization}) is assumed. We write $\chi(\dulR,t)$ in polar form, defining its phase $S(\dulR,t)$ and amplitude $a(\dulR,t)$, both real functions of $\dulR,t$. Starting from the TDSE~(\ref{eqn: tdse for chi}) and separating the real and imaginary part, two coupled equations are obtained, namely
\begin{align}
-\partial_t S(\dulR,t) &= \sum_\nu \frac{\left[\nabla_\nu S(\dulR,t)\right]^2}{2M_\nu} + \tilde{V}(\dulR,t) \label{eqn: phase eq bohmian adiabatic} \\
\partial_t a(\dulR,t) &= -\sum_\nu\frac{\left[\nabla_\nu S(\dulR,t)\right]}{M_\nu}\cdot \nabla_\nu a(\dulR,t).\label{eqn: amplitude eq bohmian adiabatic}
\end{align}
The sum over the index $\nu$ runs over the particles and the new time-dependent potential $\tilde V$ is
\begin{align}
\tilde{V}(\dulR,t) = V(\dulR) -\sum_\nu \frac{\hbar^2}{2M_\nu}\frac{\nabla_\nu^2 a(\dulR,t)}{a(\dulR,t)}.
\end{align}
The second term on the right-hand-side is the so-called quantum potential, depending on the amplitude of the wave function and therefore carrying the time-dependence of $\tilde V$. Trajectories can be used to solve Eq.~(\ref{eqn: phase eq bohmian adiabatic}), by defining the momentum as $\mathbf P_\nu(\dulR,t)=\nabla_\nu S(\dulR,t)$, i.e.
\begin{align}
\dot{\mathbf P}_\nu = -\nabla_\nu \tilde{V}(\dulR,t).\label{eqn: bohmian newton equation}
\end{align}
This Newton's equation is exactly equivalent to Eq.~(\ref{eqn: phase eq bohmian adiabatic}). 

In contrast to Eq.~(\ref{eqn: tdse for chi}), the Hamiltonian given in Eq.~(\ref{eqn: nuclear hamiltonian}) and used in the TDSE~(\ref{eqn: exact nuclear eqn}) for the evolution of the nuclear wave function of the factorization does not only contain a static Coulomb interaction (already included in the TDPES) among particles, but also a time-dependent contribution due to the electrons, that are treated fully dynamically within the exact factorization approach. Despite this difference, the above procedure, used to derive Bohmian trajectories, can be applied to Eq.~(\ref{eqn: exact nuclear eqn}). We now identify $\chi(\dulR,t)$ with the nuclear wave function in Eq.~(\ref{eqn: factorization}). Let us restrict ourselves to the case considered throughout the paper, namely with a choice of gauge that sets the vector potential to zero. The general case is considered in Refs.~\cite{mqc, long_mqc}. Then, Eq.~(\ref{eqn: phase eq bohmian adiabatic}) becomes
\begin{align}
-\partial_t S(\dulR,t) = \sum_\nu \frac{\left[\nabla_\nu S(\dulR,t)\right]^2}{2M_\nu} + \tilde{\epsilon}(\dulR,t) \label{eqn: nuclear phase eq bohmian adiabatic}
\end{align}
with
\begin{align}
\tilde{\epsilon}(\dulR,t) = \epsilon(\dulR,t)-\sum_\nu \frac{\hbar^2}{2M_\nu}\frac{\nabla_\nu^2 a(\dulR,t)}{a(\dulR,t)}\label{eqn: bohmian tdpes}
\end{align}
and $\epsilon(\dulR,t)$ given in Eq.~(\ref{eqn: tdpes}). In Eq.~(\ref{eqn: bohmian tdpes}) it is clear that the, already time-dependent, potential $\epsilon(\dulR,t)$ is corrected by the Bohmian quantum potential. From Eqs.~(\ref{eqn: nuclear phase eq bohmian adiabatic}) and~(\ref{eqn: bohmian tdpes}), Newton's equation can be derived,
\begin{align}
\dot{\mathbf P}_\nu = -\nabla_\nu \epsilon(\dulR,t)-\nabla_\nu\sum_{\nu'} \frac{\hbar^2}{2M_{\nu'}}\frac{\nabla_{\nu'}^2 a(\dulR,t)}{a(\dulR,t)}.\label{eqn: bohmian nuclear newton equation}
\end{align}
This is not the equation used in Section~\ref{sec: cl vs. qm}, as the quantum potential is not considered in the calculation of the force (see Eq.~(\ref{eqn: hamilton eom}) for comparison). It follows that it is not a priori expected that the classical trajectories evolving on $\epsilon(\dulR,t)$ alone reproduce the quantum nuclear evolution. The presence of the quantum potential in Eq.~(\ref{eqn: bohmian nuclear newton equation}) ensures that the trajectory-based treatment of the quantum dynamical problem is exactly, apart from numerical errors, equivalent to a TDSE. When this contribution is neglected, as in the results presented in Section~\ref{sec: cl vs. qm}, disagreement between exact and trajectory-based results might be observed.
Moreover, it is worth stressing that since the quantum potential is totally ignored throughout the paper, our analysis does not focus on understanding the properties of or creating a potential that allows to include quantum nuclear effects within a trajectory-based approach. We aim at describing the properties of a potential that induces effects, such as the splitting of a nuclear wave packet, related to the non-adiabatic behavior of the electrons that are coupled to the nuclei. The TDPES itself will not be able to account for purely nuclear quantum effects, such as tunnelling, which can be however expected if the quantum potential is taken into account in Eq.~(\ref{eqn: bohmian nuclear newton equation}).

\section{Step's height in ${\boldsymbol{\epsilon_{GD}(R,t)}}$}\label{app: height}
The expression of the GD part of the TDPES can be rewritten by using the relation for the phase, $\gamma_l$, of the coefficient of $C_l$, in terms of the phases, $s_l$, of $F_l(R,t)$ and the phase,
$S(R,t)$, of the nuclear wave function: $\gamma_l(R,t)=s_l(R,t)-S(R,t)$.
Together with the PNC, we
obtain then an exact expression
\begin{equation}\label{eqn: GD with phases}
\epsilon_{GD}(R,t)=\sum_{l=1,2}\left|C_l(R,t)\right|^2\dot s_l(R,t)-\dot S(R,t).
\end{equation}
The phase of the nuclear wave function appears also in the expression for the vector potential, as
\begin{equation}\label{eqn: vector potential with phases}
A(R,t) = \sum_{l=1,2}\left|C_l(R,t)\right|^2s_l'(R,t)-S'(R,t),
\end{equation}
where we neglected all terms containing the NACVs since they are negligible in the region where the steps form.

The gauge condition, $A(R,t)= 0$, can be used here to derive an expression for $\dot S(R,t)$ in Eq.~(\ref{eqn: GD with phases}). By setting Eq.~(\ref{eqn: vector potential with phases}) equal to 0, we obtain
\begin{equation}\label{eqn: S from gauge choice}
S(R,t) = \int^RdR'\sum_{l=1,2}\left|C_l(R',t)\right|^2s_l'(R',t).
\end{equation}
We insert this expression in Eq.~(\ref{eqn: GD with phases})
\begin{align}
\epsilon_{GD}(R,t)=&\sum_{l=1,2}\left|C_l(R,t)\right|^2\dot s_l(R,t) \nonumber \\
&-\int^RdR'\partial_t\sum_{l=1,2}\left|C_l(R',t)\right|^2s_l'(R',t)\\
=&\int^R dR' \partial_{R'}\sum_{l=1,2}\left|C_l(R',t)\right|^2\dot s_l(R',t)\nonumber \\
&-\int^RdR'\partial_t\sum_{l=1,2}\left|C_l(R',t)\right|^2s_l'(R',t).
\end{align}
(differentiating and integrating the first term on the right-hand-side to get the second inequality).
Throughout, we neglect any spatially-constant term in $\epsilon_{GD}(R,t)$ because it has no physical effect and we are interested in evaluating energy differences. 
After the derivatives are applied to all quantities in the sum, the remaining terms are
\begin{align}\nonumber
\epsilon_{GD}(R,t)=&\int^R dR' \sum_{l=1,2}\left[\partial_{R'}\left|C_l(R',t)\right|^2\right]\dot s_l(R',t)\\
&-\int^RdR'\sum_{l=1,2}\left[\partial_t\left|C_l(R',t)\right|^2\right]s_l'(R',t).\label{eqn: used equation for gd}
\end{align}
If the full TDSE is expanded on the adiabatic basis, neglecting the contributions from the NACVs and considering only the real part up to within terms $O(\hbar^2)$, $\dot s_l(R,t)$ can be expressed as 
\begin{equation}\label{eqn: dot phi_l}
-\dot s_l(R,t) = \frac{{s_l'}^2(R,t)}{2M} + \epsilon_{BO}^{(l)}(R).
\end{equation}
Instead, the time derivative of $|C_l(R,t)|^2$ may be written as
\begin{align}
\partial_t|C_l(R,t)|^2 = 2\frac{|C_l(R,t)|}{|\chi(R,t)|}&\Big(\partial_t|F_l(R,t)|\label{eqn: dt of C_l}\\
 &-|C_l(R,t)|\partial_t|\chi(R,t)|\Big)\nonumber
\end{align}
using the relation Eq.~(\ref{eqn: relation coefficients}).
Moreover, the time derivatives of $|F_l(R,t)|$ and $|\chi(R,t)|$ can be traded for spatial derivatives, by the equation of continuity, from the imaginary part of the TDSE expanded on the adiabatic basis,
\begin{align}
\partial_t|F_l(R,t)| &= -\frac{s_l'(R,t)}{M}\partial_R|F_l(R,t)| -\frac{s_l''(R,t)}{2M}|F_l(R,t)|,
\end{align}
and from Eq.~(\ref{eqn: exact nuclear eqn}),
\begin{equation}
\partial_t|\chi(R,t)| = -\frac{S'(R,t)}{M}\partial_R|\chi(R,t)| -\frac{S''(R,t)}{2M}|\chi(R,t)|.
\end{equation}
We now replace the explicit expressions for $S'(R,t)$
and $S''(R,t)$ from Eq.~(\ref{eqn: S from gauge choice}), and
Eq.~(\ref{eqn: dt of C_l}) becomes
\begin{align}
\partial_t|C_l(R,t)|^2 = -2\frac{s_l'(R,t)}{M} |C_l(R,t)|^2\frac{\partial_R|F_l(R,t)|}{|F_l(R,t)|}\nonumber \\
-\frac{s_l''(R,t)}{M}|C_l(R,t)|^2+|C_l(R,t)|^2\frac{S''(R,t)}{M}\nonumber\\
+2|C_l(R,t)|^2\frac{S'(R,t)}{M}\frac{\partial_R|\chi(R,t)|}{|\chi(R,t)|},
\end{align}
 where
\begin{align}
\frac{\partial_R|\chi(R,t)|}{|\chi(R,t)|}&=\frac{\sum_{l=1,2}|F_l(R,t)|\partial_R|F_l(R,t)|}{|\chi(R,t)|^2}\nonumber \\
&= \sum_{l=1,2}|C_l(R,t)|^2\frac{\partial_R|F_l(R,t)|}{|F_l(R,t)|},
\end{align}
which follows from Eq.~(\ref{eqn: chi and Fl}). These expressions are used in Eq.~(\ref{eqn: used equation for gd}), and, after some algebra, $\epsilon_{GD}(R,t)$ becomes
\begin{align}
\epsilon_{GD}&(R,t) = \nonumber \\
&\int^R dR' \left(\mathcal I^{(BO)}(R',t)+\mathcal I^{(I)}(R',t)+\mathcal I^{(II)}(R',t)\right)\label{eqn: exact gd}
\end{align}
The three terms in the integral are the $BO$ term
\begin{equation}\label{eqn: leading term in gd}
\mathcal I^{(BO)}=\left(\epsilon_{BO}^{(2)}-\epsilon_{BO}^{(1)}\right)\partial_{R}|C_1|^2
\end{equation}
and two small corrections labeled $(I)$ (first)
\begin{equation}\label{eqn: remaining term in gd}
\mathcal I^{(I)}=\frac{\left[s_2'-s_1'\right]^2}{M}\left[\frac{\partial_R|F_1|}{|F_1|}+\frac{\partial_R|F_2|}{|F_2|}\right]|C_1|^2|C_2|^2
\end{equation}
and $(II)$ (second)
\begin{equation}\label{eqn: second derivative term in gd}
\mathcal I^{(II)}=\frac{\left[s_1's_1''+s_2's_2''-s_1''s_2'-s_1's_2''\right]^2}{M}|C_1|^2|C_2|^2.
\end{equation}
So far, the only approximation has been neglecting the term $\mathcal
O(\hbar^2)$ in the evolution equation~(\ref{eqn: dot phi_l}) for $\dot
s_l(R,t)$. Note that the PNC, in the form
$\partial_R|C_2(R,t)|^2=-\partial_R|C_1(R,t)|^2$, has been used in
Eq.~(\ref{eqn: leading term in gd}). The left panel in Fig.~\ref{fig:
  terms gd height} show for the times indicated in the
plots, the functions in Eqs.~(\ref{eqn: leading term in
  gd}),~(\ref{eqn: remaining term in gd}) and~(\ref{eqn: second
  derivative term in gd}). It is clear
that the $BO$ term is by far the dominant contribution to
$\epsilon_{GD}(R,t)$ and that the contribution from Eq.~(\ref{eqn: second derivative term in gd}) can be neglected.
\begin{figure}[h!]
 \begin{center}
 \includegraphics[angle=270,width=.85\textwidth]{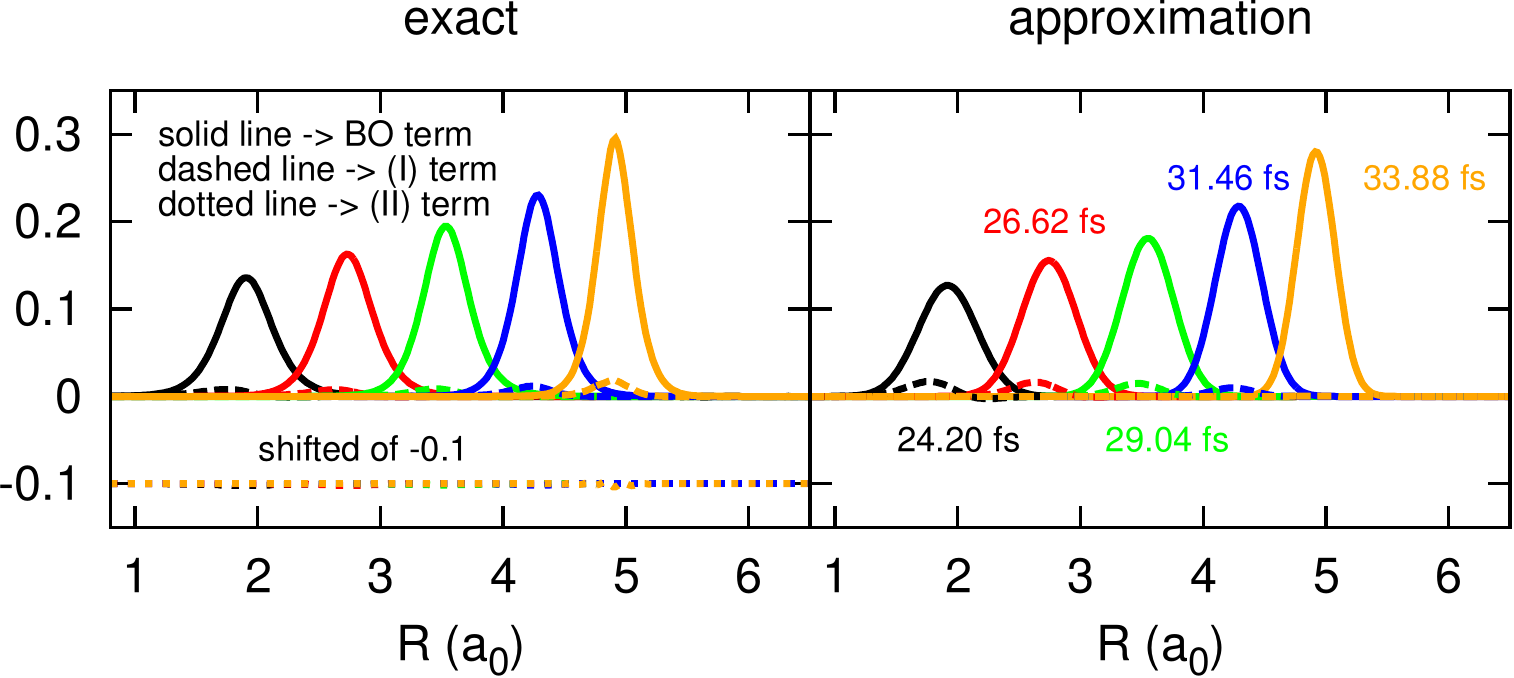}
 \caption{Left: comparison between Eq.~(\ref{eqn: leading term in gd}), solid lines, Eq.~(\ref{eqn: remaining term in gd}), dashed lines, and Eq.~(\ref{eqn: second derivative term in gd}), dotted lines, at some times after the splitting of the nuclear wave packet. The figure shows that the average of the BO energies in the region of the step, weighted by $\partial_R|C_1(R,t)|^2$ is indeed the leading term. Right: approximations to Eqs.~(\ref{eqn: leading term in gd}) and~(\ref{eqn: remaining term in gd}) based on quasiclassical arguments.}
 \label{fig: terms gd height}
 \end{center}
\end{figure}
The right panel in Fig.~\ref{fig: terms gd height} show an approximation to Eqs.~(\ref{eqn: leading term in gd}) and~(\ref{eqn: remaining term in gd}). To have an analytic expression for $\mathcal{I}^{(BO)}$, first we make use of the error-function structure of Eq.~(\ref{eqn: fitting function}). This means its spatial derivative  is a Gaussian centered at $R_0$ with variance $\alpha^{-2}(t)$, namely
\begin{align}\label{eqn: dr C}
\partial_R\left|C_1(R,t)\right|^2 = \frac{\alpha(t)}{\sqrt{\pi}}e^{-\alpha^2(t)(R-R_0(t))^2}.
\end{align}
Inserted into Eq.~(\ref{eqn: leading term in gd}), this gives an expression for the dominant contribution to $\epsilon_{GD}(R,t)$ in terms of the BOPESs, the crossover point $R_0$ and width $\alpha(t)$. 
For the correction term Eq.~(\ref{eqn: remaining term in gd}), we use Eq.~(\ref{eqn: dot phi_l}) to identify the spatial derivative of the phase $s_l(R,t)$, as the momentum associated to the motion of the wave packet on the $l$-th BO surface and we approximate $|F_l(R,t)|^2$, with a Gaussian centered in $R_l^{\textrm{qm}}(t)$ with variance $\sigma_l^2(t)$. Therefore, we use the following expression
\begin{align}
\frac{\partial_R|F_l(R,t)|}{|F_l(R,t)|} &= -\frac{R-R_l^{\textrm{qm}}(t)}{\sigma_l^2(t)},
\end{align}
where $R_l^{\textrm{qm}}$ and $\sigma_l^2$ are obtained from exact calculations, according to
\begin{equation}\label{eqn: l-th mean nuclear position}
 R_l^{\textrm{qm}}(t) = \frac{\int dR\,R|F_l(R,t)|^2}{\int dR|F_l(R,t)|^2}
\end{equation}
and
\begin{equation}\label{eqn: variance from Fl}
 \sigma_l^2(t) = \frac{2\int dR\, \left[R-R_l^{\textrm{qm}}(t)\right]^2|F_l(R,t)|^2}{\int dR |F_l(R,t)|^2}.
\end{equation}
The height $2h(t)$ of the steps can be estimated as the energy difference between two points $R^+>R_0$ and $R^-<R_0$, chosen far enough from $R_0$, such that it guarantees that $\epsilon_{GD}(R>R^+,t)$ and $\epsilon_{GD}(R<R^-,t)$ are constant. Therefore,
\begin{align}
2h(t)&=\epsilon_{GD}(R^+,t)-\epsilon_{GD}(R^-,t) \nonumber \\
&=\int^{R^+}dR'\left[\cdots\right]-\int^{R^-}dR'\left[\cdots\right]
\end{align}
where the dots in square brackets represent the function under the integral sign in Eq.~(\ref{eqn: exact gd}). Since $R^+>R^-$, we can split the first integral in two parts 
\begin{equation}
2h(t)=\int^{R^-}dR'\left[\cdots\right]+\int_{R^-}^{R^+}dR'\left[\cdots\right]-\int^{R^-}dR'\left[\cdots\right]
\end{equation}
and the remaining term is only the integral performed over the region from $R^-$ to $R^+$. Since the functions in the integral are localized around $R_0$ and rapidly decay to zero (the Gaussian in Eq.~(\ref{eqn: leading term in gd}) and the product $\left|C_1(R,t)\right|^2\left|C_2(R,t)\right|^2$ in Eq.~(\ref{eqn: remaining term in gd})), the boundaries of the integral can be set to infinity. The final result, as shown in Eq.~(\ref{eqn: leading term bis}), is
\begin{align}
2h(t)\simeq\frac{\alpha}{\sqrt{\pi}}\int_{-\infty}^{+\infty} dR\,\Delta_{21}^{(BO)}(R)\, e^{-\alpha^2(t)(R-R_0(t))^2}
\end{align}
with $\Delta_{21}^{(BO)}(R)=\epsilon_{BO}^{(2)}(R)-\epsilon_{BO}^{(1)}(R)$, with correction
\begin{align}
\mathcal C(t)=-\frac{\left[P_2(t)-P_1(t)\right]^2}{4M}\int_{-\infty}^{+\infty} dR\left|C_1(R,t)\right|^2\left|C_2(R,t)\right|^2 \nonumber \\
\left[\frac{R-R_1^{\textrm{qm}}(t)}{\sigma_1^2(t)}+\frac{R-R_2^{\textrm{qm}}(t)}{\sigma_2^2(t)}\right].
\end{align}


%

\end{document}